\documentclass[sigconf]{acmart}
\AtBeginDocument{%
  \providecommand\BibTeX{{%
    \normalfont B\kern-0.5em{\scshape i\kern-0.25em b}\kern-0.8em\TeX}}}

\copyrightyear{2024}
\acmYear{2024}
\setcopyright{acmlicensed}\acmConference[KDD '24]{Proceedings of the 30th
ACM SIGKDD Conference on Knowledge Discovery and Data Mining}{August
25--29, 2024}{Barcelona, Spain}
\acmBooktitle{Proceedings of the 30th ACM SIGKDD Conference on Knowledge
Discovery and Data Mining (KDD '24), August 25--29, 2024, Barcelona, Spain}
\acmDOI{10.1145/3637528.3671882}
\acmISBN{979-8-4007-0490-1/24/08}

\usepackage[utf8]{inputenc} 
\usepackage[T1]{fontenc}    
\usepackage{hyperref}       
\usepackage{amsfonts}       
\usepackage{graphicx}
\usepackage{amsthm}
\usepackage{amsmath}
\usepackage{subfigure}
\usepackage{multirow}
\usepackage{enumitem}
\usepackage{bbding}
\usepackage{colortbl}
\usepackage{arydshln}

\definecolor{RPD_pos0}{RGB}{249, 218, 183} 
\definecolor{RPD_pos1}{RGB}{253, 205, 170}
\definecolor{RPD_pos2}{RGB}{255, 192, 150} 
\definecolor{RPD_neg0}{RGB}{150, 201, 232} 
\definecolor{RPD_neg1}{RGB}{119, 187, 230}
\definecolor{RPD_neg2}{RGB}{115, 180, 234}

\newcommand{\paratitle}[1]{\vspace{0.8ex}\noindent\textbf{#1}}

\begin{document}

\title{Neural Retrievers are Biased Towards LLM-Generated Content}

\author{Sunhao Dai}
\author{Yuqi Zhou}
\affiliation{
  \institution{\mbox{Gaoling School of Artificial Intelligence}\\Renmin University of China}
\city{Beijing}
  \country{China}
  }
\email{{sunhaodai, yuqizhou}@ruc.edu.cn}

\author{Liang Pang}
\affiliation{%
  \institution{CAS Key Laboratory of AI Safety\\Institute of Computing Technology Chinese Academy of Sciences}
  \city{Beijing}
  \country{China}
}
\email{pangliang@ict.ac.cn}

\author{Weihao Liu}
\author{Xiaolin Hu}
\affiliation{%
  \institution{\mbox{Gaoling School of Artificial Intelligence}\\Renmin University of China}
\city{Beijing}
  \country{China}
}
\email{{weihaoliu, xiaolinhu}@ruc.edu.cn}

\author{Yong Liu}
\author{Xiao Zhang}
\affiliation{%
  \institution{\mbox{Gaoling School of Artificial Intelligence}\\Renmin University of China}
\city{Beijing}
  \country{China}
}
\email{{liuyonggsai, zhangx89}@ruc.edu.cn}

\author{Gang Wang}
\affiliation{%
 \institution{Huawei Noah's Ark Lab}
 \city{Shenzhen}
  \country{China}
}
\email{wanggang110@huawei.com}

\author{Jun Xu}
\authornote{Jun Xu is the corresponding author. Work partially done at Engineering Research Center of Next-Generation Intelligent Search and Recommendation, Ministry of Education.}
\affiliation{%
  \institution{\mbox{Gaoling School of Artificial Intelligence}\\Renmin University of China}
 \city{Beijing}
  \country{China}
}
\email{junxu@ruc.edu.cn}

\renewcommand{\authors}{Sunhao Dai, Yuqi Zhou, Liang Pang, Weihao Liu, Xiaolin Hu, Yong Liu, Xiao Zhang, Gang Wang and Jun Xu}
\renewcommand{\shortauthors}{Sunhao Dai et al.}

\begin{abstract}

Recently, the emergence of large language models (LLMs) has revolutionized the paradigm of information retrieval (IR) applications, especially in web search, by generating vast amounts of human-like texts on the Internet. As a result, IR systems in the LLM era are facing a new challenge: the indexed documents are now not only written by human beings but also automatically generated by the LLMs. How these LLM-generated documents influence the IR systems is a pressing and still unexplored question. In this work, we conduct a quantitative evaluation of IR models in scenarios where both human-written and LLM-generated texts are involved. Surprisingly, our findings indicate that neural retrieval models tend to rank LLM-generated documents higher. We refer to this category of biases in neural retrievers towards the LLM-generated content as the \textbf{source bias}. Moreover, we discover that this bias is not confined to the first-stage neural retrievers, but extends to the second-stage neural re-rankers. Then, in-depth analyses from the perspective of text compression indicate that LLM-generated texts exhibit more focused semantics with less noise, making it easier for neural retrieval models to semantic match. To mitigate the source bias, we also propose a plug-and-play debiased constraint for the optimization objective, and experimental results show its effectiveness. Finally, we discuss the potential severe concerns stemming from the observed source bias and hope our findings can serve as a critical wake-up call to the IR community and beyond. To facilitate future explorations of IR in the LLM era, the constructed two new benchmarks are available at \url{https://github.com/KID-22/Source-Bias}.

\end{abstract}




\maketitle

\section{Introduction}

\begin{figure}[t]
  \centering
  \subfigcapskip=-1pt
  \subfigure[IR in the Pre-LLM Era]
    {
    \includegraphics[width=0.90\columnwidth]{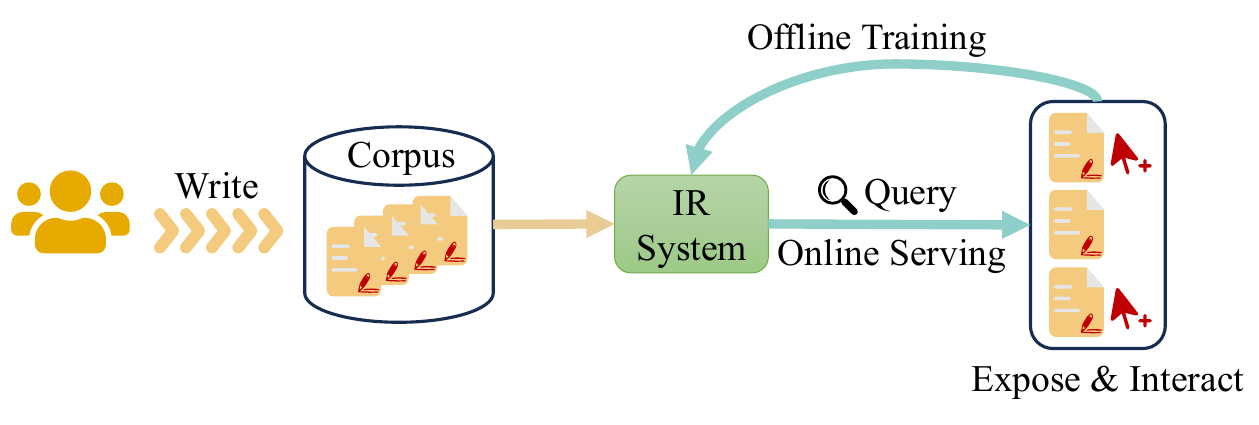}
    \label{fig:intro_era_0}
   }
  \subfigure[IR in the LLM Era]{
    \includegraphics[width=0.90\columnwidth]{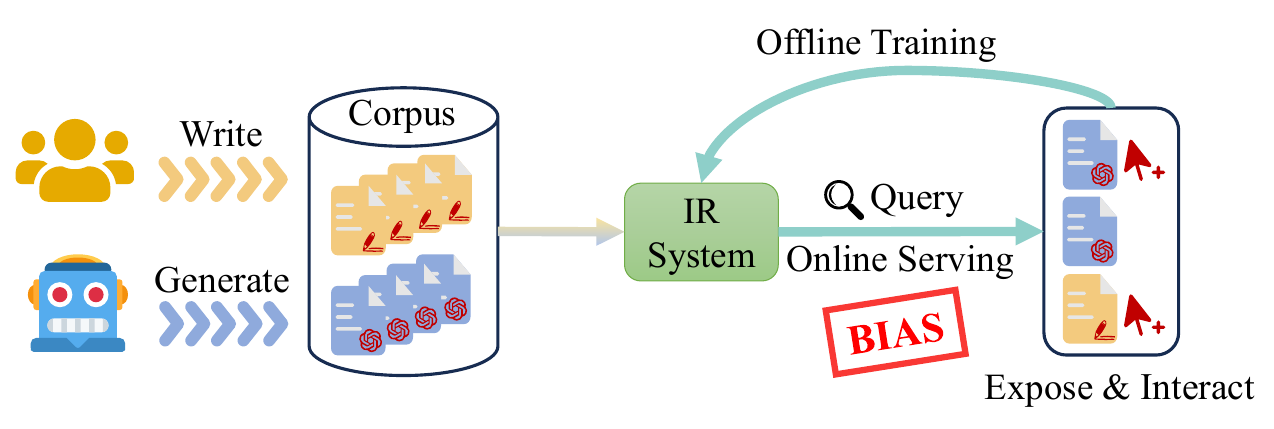}
    \label{fig:intro_era_1}
  }
    \vspace{-0.15in}
    \caption{The overview evolution of IR paradigm from the Pre-LLM era to the LLM era.}
    \label{fig:intro_era}  
\end{figure}

With the advent of large language models (LLMs), exemplified by ChatGPT, the field of artificial intelligence generated content (AIGC) has surged to new heights of prosperity~\cite{cao2023comprehensive, wu2023ai}. 
LLMs have demonstrated their remarkable capabilities in automatically generating human-like text at scale, resulting in the Internet being inundated with an unprecedented volume of AIGC content~\cite{wei2022emergent, spitale2023ai}. 
This influx of LLM-generated content has fundamentally reshaped the digital ecosystem, challenging conventional paradigms of content creation, dissemination, and information access on the Internet~\cite{ai2023information, zhu2023large}.

Meanwhile, information retrieval (IR) systems have become indispensable for navigating and accessing the Internet's vast information landscape~\cite{singhal2001modern, manning2009introduction}.
As illustrated in~\autoref{fig:intro_era}, in the era preceding the widespread emergence of LLMs, IR systems focused on retrieving documents solely from the human-written corpus in response to users' queries~\cite{liu2009learning, li2022learning, xu2007adarank}.
However, the proliferation of AIGC driven by LLMs has expanded the corpus of IR systems to include both human-written and LLM-generated texts. 
Consequently, this paradigm shift raises a fundamental research question: \textbf{What is the impact of the proliferation of generated content on IR systems?} 
We aim to explore whether existing retrieval models tend to prioritize LLM-generated text over human-written text, even when both convey similar semantic information.
If this holds, LLMs may dominate information access, particularly as their generated content is rapidly growing on the Internet~\cite{hanley2023machine}.

To approach the fundamental research question, we decompose it into four specific research questions. The first question is \textbf{RQ1: How to construct an environment to evaluate IR models in the LLM era?} 
Given the lack of public retrieval benchmarks encompassing both human-written and LLM-generated texts, we propose an innovative and practical method to create such a realistic evaluation environment without the need of costly human annotation. Specifically, we leverage the original human-written texts as the instruction conditions to prompt LLMs to generate rewritten text copies while preserving the same semantic meaning. 
In this way, we can confidently assign the same relevancy labels to LLM-generated data according to the original labels. Extensive empirical analysis validates the quality of our constructed environment, demonstrating its effectiveness in mirroring real-world IR scenarios in the LLM era. As a result, we introduce two new benchmarks, SciFact+AIGC and NQ320K+AIGC, tailored for IR research in the LLM era.

\begin{figure*}[t]  
    \centering    
    \includegraphics[width=1\linewidth]{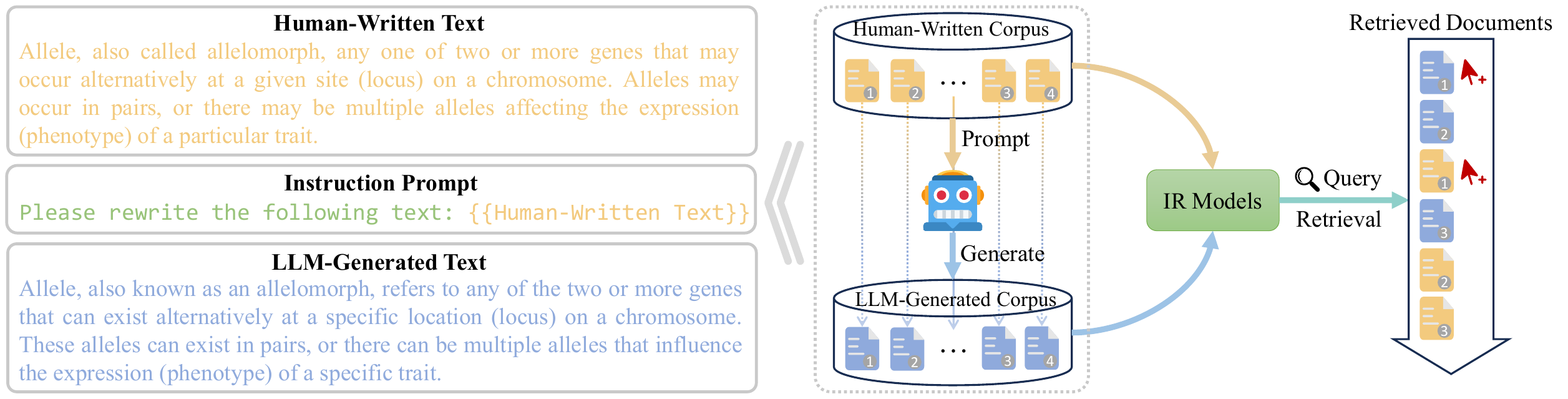}
    \caption{The overall paradigm of the proposed evaluation framework for IR in the LLM era.}
    \label{fig:evaluation_framework}  
\end{figure*}

With the constructed environment, we further explore \textbf{RQ2: Are retrieval models biased towards LLM-generated content?} We conduct comprehensive experiments with various representative retrieval models, ranging from traditional lexical models to modern neural models based on pretrained language models (PLMs)~\cite{guo2020deep, zhao2022dense, yates2021pretrained, guo2022semantic}. Surprisingly, we uncover that neural retrievers are biased towards LLM-generated texts, i.e., tend to rank LLM-generated texts in higher positions.
We refer to this as \textbf{source bias}, as the neural retrievers favor content from specific sources (i.e., LLM-generated content).
Further experiments indicate that the source bias not only extends to the second-stage neural re-rankers from the first-stage retrieval but also manifests more severely.
These findings corroborate the prevalence of source bias in neural retrieval models.

Then, what we are curious about is \textbf{RQ3: Why are neural retrieval models biased towards LLM-generated texts?}
Inspired by the recent studies positing LLMs as lossless compressors~\cite{deletang2023language}, we analyze the cause of source bias from the viewpoint of text compression. 
Our analysis of singular values~\cite{klema1980singular} in different corpora reveals that LLM-generated texts exhibit more focused semantics with minimal noise, enhancing their suitability for semantic matching. 
Furthermore, our in-depth perplexity analysis shows that LLM-generated texts consistently achieve lower perplexity scores, which indicates a higher degree of comprehensibility and confidence from the PLM's perspective. 
These observations collectively suggest that LLM-generated texts are more readily understandable to PLM-based neural retrievers, thereby resulting in source bias.

Finally, we try to answer
\textbf{RQ4: How to mitigate source bias in neural retrieval models?} To tackle this, we propose an intuitive yet effective debiased constraint.
This constraint is designed to penalize biased samples during the optimization process, thereby shifting the focus of retrieval models from exploiting inherent shortcuts to emphasizing semantic relevance.
Besides, our debiased constraint is model-agnostic and can be plugged and played to the ranking optimization objectives of various neural retrieval models.
Furthermore, it offers the capability to control the degree of bias removal, offering the flexibility to balance the treatment between the two sources of content based on specific requirements and environmental considerations.

Last but not least, we discuss the potential emerging concerns stemming from source bias, highlighting the risk of human-written content being gradually inaccessible, especially due to the rapidly increasing LLM-generated content on the Internet~\cite{hanley2023machine, bengio2023managing}. Furthermore, source bias could be maliciously exploited to manipulate algorithms and potentially amplify the spread of misinformation, posing a threat to online security.
In light of these pressing issues, we hope that our findings serve as a resounding wake-up call to all stakeholders involved in IR systems and beyond.

In summary, the contributions of this paper are as follows:  

(1) We introduce a more realistic paradigm of IR systems considering the growing prosperity of AIGC, where the retrieval corpus consists of both human-written and LLM-generated texts. We then uncover a new inherent bias in both neural retrieves and re-rankers preferring LLM-generated content, termed as source bias.

(2) We provide an in-depth analysis and insights of source bias from a text compression perspective, which indicates that LLM-generated texts maintain more focused semantics with minimal noise and are more readily comprehensible for neural retrievers.

(3) We propose a debiased constraint to penalize the biased samples during optimization, and experimental results demonstrate its effectiveness in mitigating source bias in different degrees.

(4) We also provide two new benchmarks, SciFact+AIGC and NQ320K+AIGC, which contain both high-quality human-written and various LLM-generated corpus and corresponding relevancy labels. We believe these two benchmarks can serve as valuable resources for facilitating future research of IR in the LLM era.

\section{RQ1: Environment Construction}
With the increasing usage of LLMs in generating texts (e.g., paraphrasing or rewriting), the corpus of IR systems includes both human-written and LLM-generated texts nowadays.
Constructing an IR dataset in the LLM era typically involves two steps: collecting both human-written and LLM-generated corpora and then employing human evaluators to annotate relevancy labels for each query-document pair. Given that LLM-generated content is currently unidentifiable~\cite{Sadasivan2023CanAT} and the significant cost of human annotation, we introduce a natural and practical framework for quantitatively evaluating retrieval models in the LLM era, as shown in~\autoref{fig:evaluation_framework}.

To better align with real-world scenarios, the evaluation environments should meet the following three essential criteria.
\textbf{Firstly}, it is imperative to distinguish between human-written and LLM-generated texts within the corpus.
\textbf{Secondly}, we need access to relevancy labels for LLM-generated data in response to queries. 
\textbf{Thirdly}, each human-written text should better have a corresponding LLM-generated counterpart with the same semantics, ensuring the most effective and fair evaluation.

\begin{table*}
\centering
\caption{Statistics of the constructed two datasets. Avg. Doc / Query means the average number of relevant documents per query.}
\vspace{-0.2cm}
\label{tab:stats_datasets}
\resizebox{1\linewidth}{!}{
\begin{tabular}{ccccccccc}
\hline \hline
\multirow{2}{*}{Dataset}    & \multirow{2}{*}{\# Test Queries} & \multirow{2}{*}{\# Avg. Query Length} & \multicolumn{3}{c}{Human-Written Corpus} & \multicolumn{3}{c}{Llama2-Generated Corpus ; ChatGPT-Generated Corpus}\\
\cmidrule(lr){4-6} \cmidrule(lr){7-9}
           &          &           & \# Corpus        &   Avg. Doc Length      & Avg. Doc / Query  & \# Corpus     &   Avg. Doc Length      & Avg. Doc / Query\\
\hline
SciFact+AIGC  & 300      & 12.38      & 5,183            & 201.81  & 1.1    & 5,183 ; 5,183  & 192.66 ; 203.57     & 1.1 ; 1.1        \\
NQ320K+AIGC &  7,830     &  9.24      & 109,739         & 199.79  & 1.0  & 109,739 ; --  & 174.49 ; -- &  1.0 ; --      \\
\hline \hline
\end{tabular}}
\end{table*}

\subsection{Notation}
Formally, in the Pre-LLM era, given a query $q \in \mathcal{Q}$ where $\mathcal{Q}$ is the set of all queries, the traditional IR system aims to retrieve a list of top-$K$ relevant documents $\{d^{(1)}, d^{(2)}, \ldots , d^{(K)}\}$ from a corpus $\mathcal{C}^H = \{d_1^H, d_2^H, \ldots d_{N}^H\}$ which consists of $N$ human-written documents.  
However, in the era of LLMs, there is also LLM-generated text in the corpus. 
To evaluate the IR models in the LLM era, we also create an additional LLM-generated corpus $\mathcal{C}^G = \{d_1^G, d_2^G, \ldots, d_N^G\}$ where each document is generated by a LLM, e.g., $d_1^G$ can be created by prompting ChatGPT to rewrite $d_1^H$ while preserving its original semantics information. Consequently, given a query $q$, the objective of a retriever in the LLM era is to return the top-$K$ relevant documents from the mixed corpus $\mathcal{C} = \mathcal{C}^H \bigcup \mathcal{C}^G$.

\subsection{Constructing IR Datasets in the LLM Era} \label{sec: data_gen}

In this section, we prompt LLMs to rewrite human-written corpus to build two new standard retrieval datasets: SciFact+AIGC and NQ320K+AIGC. 
These two new datasets can serve as valuable resources to facilitate future research of IR in the LLM era.

\begin{table*}
\caption{Performance comparison of retrieval models on the sole human-written or Llama2-generated corpus on SciFact+AIGC and NQ320K+AIGC datasets. For brevity, we omit the percent sign `$\%$' of ranking metrics in subsequent tables and figures. }
\centering
\vspace{-0.2cm}
\label{tab:quality2_llama2_gaps}
\resizebox{1.0\textwidth}{!}{
\begin{tabular}{ccccccccccccccc}
\hline\hline
Model     & \multirow{2}{*}{Model}       & \multirow{2}{*}{Corpus}                         & \multicolumn{6}{c}{SciFact+AIGC} &         \multicolumn{6}{c}{NQ320K+AIGC}       \\
\cmidrule(lr){4-9} \cmidrule(lr){10-15}
Type      &             &                                & NDCG@1   & NDCG@3 & NDCG@5 & MAP@1 & MAP@3 & MAP@5 & NDCG@1  & NDCG@3 & NDCG@5 & MAP@1 & MAP@3 & MAP@5  \\
\hline
\multirow[m]{4}{*}{Lexical} & \multirow[m]{2}{*}{TF-IDF}     & Human-Written & 42.0     & 49.5   & 52.7   & 40.7  & 47.1  & 49.0  & 12.2    & 15.8   & 16.8   & 12.2  & 14.9  & 15.5   \\
        &            & LLM-Generated & 43.0     & 49.8   & 52.6   & 40.8  & 47.5  & 49.2  & 9.4     & 12.6   & 13.9   & 9.4   & 11.8  & 12.5   \\
\cline{2-15}        
        & \multirow[m]{2}{*}{BM25}       & Human-Written & 46.0     & 54.2   & 56.3   & 43.8  & 51.5  & 52.8  & 12.9    & 16.3   & 17.6   & 12.9  & 15.5  & 16.2   \\
        &            & LLM-Generated & 46.3     & 53.6   & 55.3   & 44.1  & 51.1  & 52.2  & 11.9    & 15.3   & 16.5   & 11.9  & 14.5  & 15.1   \\
\hline        
\multirow[m]{8}{*}{Neural}  & \multirow[m]{2}{*}{ANCE}       & Human-Written & 38.7     & 44.3   & 46.5   & 36.3  & 41.9  & 43.3  & 50.6    & 60.0   & 62.2   & 50.6  & 57.7  & 58.9   \\
        &            & LLM-Generated & 41.0     & 46.0   & 48.2   & 37.8  & 43.5  & 45.0  & 49.3    & 58.8   & 61.2   & 49.3  & 56.5  & 57.8   \\
\cline{2-15}        
        & \multirow[m]{2}{*}{BERM}       & Human-Written & 37.0     & 42.1   & 44.2   & 34.7  & 39.7  & 41.3  & 49.2    & 58.3   & 60.4   & 49.2  & 56.1  & 57.3   \\
        &            & LLM-Generated & 40.7     & 44.5   & 46.2   & 37.7  & 42.3  & 43.5  & 48.4    & 57.5   & 59.8   & 48.4  & 55.3  & 56.5   \\
\cline{2-15}        
        & \multirow[m]{2}{*}{TAS-B}      & Human-Written & 52.7     & 58.1   & 60.2   & 49.9  & 55.6  & 57.2  & 53.4    & 63.0   & 65.4   & 53.4  & 60.7  & 62.0   \\
        &            & LLM-Generated & 50.7     & 57.0   & 58.9   & 48.0  & 54.6  & 55.9  & 51.9    & 62.3   & 64.7   & 51.9  & 59.8  & 61.1   \\
\cline{2-15}        
        & \multirow[m]{2}{*}{Contriever} & Human-Written & 54.0     & 61.8   & 63.2   & 51.4  & 58.9  & 60.0  & 58.2    & 68.4   & 70.3   & 58.2  & 65.9  & 67.0   \\
        &            & LLM-Generated & 55.7     & 62.0   & 64.8   & 52.9  & 59.5  & 61.5  & 57.1    & 67.5   & 69.8   & 57.1  & 64.9  & 66.2  \\
\hline\hline
\end{tabular}
}
\end{table*}

\subsubsection{Human-Written Corpus} 
We first choose two widely used retrieval datasets written by humans in the Pre-LLM era as the seed data: SciFact and NQ320K. SciFact\footnote{\url{https://allenai.org/data/scifact}}~\cite{wadden2020fact} dataset aims to retrieve evidence from the research literature containing scientific paper abstracts for fact-checking. NQ320K\footnote{\url{https://ai.google.com/research/NaturalQuestions}}~\cite{kwiatkowski2019natural} is based on the Natural Questions (NQ) dataset from Google, where the documents are gathered from Wikipedia pages, and the queries are natural language questions. Following the practice in BEIR benchmark~\cite{thakur2021beir}, we process these two datasets in a standard format: corpus $\mathcal{C}^H$, queries $\mathcal{Q}$, and relevancy labels $\mathcal{R}^{H} = \{(q_m,d^{H}_m,r_m)\}^M_{m=1}$, where $M$ is the number of labeled query-document pairs in the dataset.

\subsubsection{LLM-Generated Corpus} 
For the LLM-generated corpus, we repurpose the original human-written corpus as our seed data and instruct LLMs to rewrite each given text from the human-written corpus. As the written text generated by LLM carries almost the same semantic information as the original human-written text, we can assign the same relevancy labels to new <query, LLM-generated document> pairs as those assigned to the original labeled <query, human-written document> pairs.

Our instruction is straightforward: ``\textit{Please rewrite the following text: \{\{human-written text\}\}}'', as illustrated in the left part of~\autoref{fig:evaluation_framework}. This straightforward instruction enables LLMs to generate text without too many constraints while maintaining semantic equivalence to the original human-written text. Specifically, we choose Llama2~\cite{touvron2023llama} and ChatGPT to rewrite each seed human-written corpus, as Llama2 and ChatGPT are both the most widely-used and nearly the state-of-the-art open-sourced and closed-source LLM, respectively. We only generate texts with ChatGPT corresponding to the texts in SciFact dataset, mainly due to the significant cost involved in processing the larger NQ320K dataset.

For the LLM-generated corpus, we conduct post-processing to remove unrelated parts of the original response from LLM like ``Sure, here's a possible rewrite of the text:''. As a result, we can obtain two corresponding LLM-generated corpora with SciFact and NQ320K as seed data. After that, we extend the original labels of query and human-written text $\mathcal{R}^{H} = \{(q_m,d^{H}_m,r_m)\}^M_{m=1}$ to get the corresponding label of LLM-generated text $\mathcal{R}^{G} = \{(q_m,d^{G}_m,r_m)\}^M_{m=1}$. We will validate the quality of the datasets in the following section. Combining each original human-written corpus $\mathcal{C}^{H}$ with its corresponding LLM-generated corpus $\mathcal{C}^{G}$, original queries $\mathcal{Q}$, and labels $\mathcal{R}^{H} \bigcup \mathcal{R}^{G}$, we can create two new datasets, denoted as SciFact+AIGC and NQ320K+AIGC. ~\autoref{tab:stats_datasets} summarizes the statistics of the proposed two datasets.

\begin{figure}[t]
  \subfigure[Jaccard Similarity]
    {
    \includegraphics[width=0.46\columnwidth]{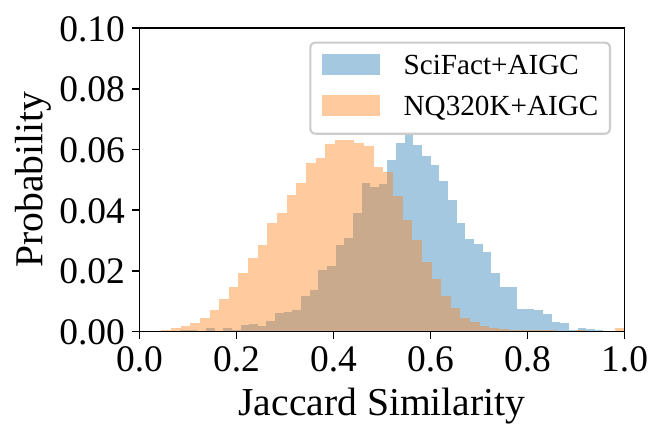}
    \label{fig:scifact_jaccard}
   }
   \quad
  \hspace{-0.25in}
  \subfigure[Overlap]{
    \includegraphics[width=0.46\columnwidth]{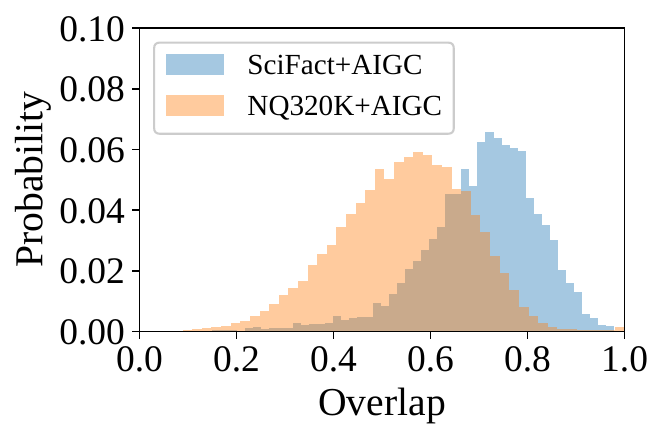}
    \label{fig:nfcorpus_jaccard}
  }
    \vspace{-0.3cm}
    \caption{Distribution of term Jaccard similarity and overlap between Llama2-generated and human-written corpora.}
    \label{fig:analysis_of_jaccard_overlap}
\end{figure}

\subsection{Statistics and Quality Validation of Datasets}
Take the Llama2-generated data as an example, we conduct the statistics and quality validation of the constructed datasets. The analysis of ChatGPT-generated datasets shows similar observations and conclusions and is omitted due to the page limitation.

\subsubsection{Term-based Statistics and Analysis}
We first analyze the term-based similarity between the LLM-generated corpus and the human-written corpus. Specifically, we compute the Jaccard similarity  ($\frac{|d^G \bigcap d^H|}{|d^G \bigcup d^H|}$) and the overlap ($\frac{|d^G \bigcap d^H|}{|d^H|}$) between each LLM-generated document and orginal human-written document. 
As shown in~\autoref{fig:analysis_of_jaccard_overlap}, both the Jaccard similarity and overlap distributions exhibit normal distribution, with peaks at about 0.6 and 0.8 for SciFact+AIGC, and about 0.4 and 0.6 for NQ320K+AIGC, respectively. These observations suggest that while there is a considerable overlap of terms between the LLM-generated text and the original human-written text, there are also distinct differences, especially noticeable in the NQ320K+AIGC dataset.

\begin{figure}[t]
  \subfigure[SciFact+AIGC]
    {
    \includegraphics[width=0.45\columnwidth]{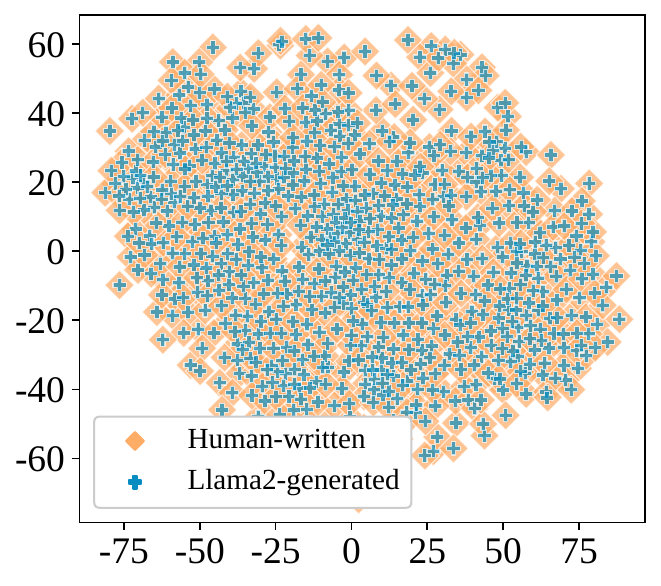}
    \label{fig:scifact_t_sne}
   }
   \quad
  \hspace{-0.25in}
  \subfigure[NQ320K+AIGC]{
    \includegraphics[width=0.45\columnwidth]{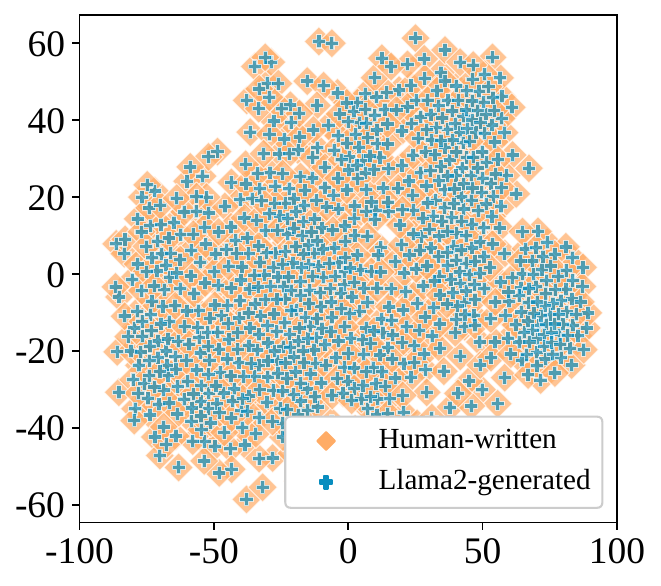}
    \label{fig:nq320k_t_sne}
  }
    \vspace{-0.3cm}
    \caption{Semantic embedding visualization of different corpora on SciFact+AIGC and NQ320K+AIGC datasets.}
    \label{fig:analysis_of_embedding_t_sne}
\end{figure}

\begin{figure}[t]  
    \centering    
    \includegraphics[width=0.85\linewidth]{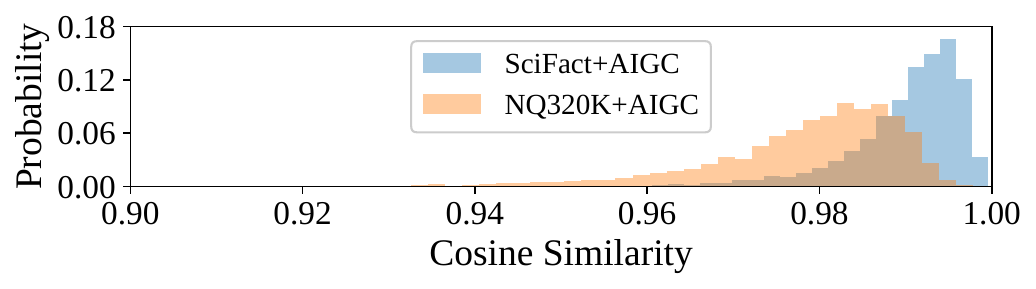}
    \vspace{-0.3cm}
    \caption{Distribution of cosine similarity of semantic embedding between Llama2-generated and human-written corpora.}
    \label{fig:analysis_of_embedding_cosine}  
\end{figure}

\subsubsection{Semantic-based Statistics and Analysis}
For the LLM-generated texts, a pivotal consideration is whether they faithfully preserve the underlying semantics of the corresponding human-written corpus. If they indeed do so, we then can confidently assign them the same relevancy labels as the labels of their corresponding original human-written texts given each query.

To assess this, we first leverage the OpenAI embedding model\footnote{text-embedding-ada-002:\url{https://platform.openai.com/docs/guides/embeddings}} to acquire semantic embeddings for both the LLM-generated and human-written texts. Subsequently, we visualize these embeddings through T-SNE~\cite{van2008visualizing} in \autoref{fig:analysis_of_embedding_t_sne}. We observe a strikingly close overlap between the Llama2-generated corpus and the human-written corpus in the latent space. This observation strongly suggests that these LLM-generated corpora adeptly preserve the original semantics. Moreover, we delve into the cosine similarity of semantic embeddings between the LLM-generated text and their corresponding human-written counterparts. The results, as shown in~\autoref{fig:analysis_of_embedding_cosine}, also indicate a high degree of similarity, with most values exceeding 0.95, affirming the faithful preservation of semantics in LLM-generated text. Hence, for each query-document pair $(q, d^G)$, we can confidently assign the relevancy label $r$ to be the same as that of $(q, d^H)$.

\begin{table}[t]
\centering
\caption{Verification of semantics and text quality with human evaluation.
The numbers in parentheses represent the proportion agreed upon by all three human annotators for each option.}
\label{tab: human_eval}
\resizebox{1 \columnwidth}{!}{
\begin{tabular}{cccccc}
\toprule
      \multicolumn{3}{c}{SciFact+AIGC} &         \multicolumn{3}{c}{NQ320K+AIGC}        \\
\midrule
\multicolumn{6}{c}{Which document is more relevant to the given query?} \\
Human & LLM & Equal & Human & LLM & Equal  \\
$0.0 \% (0.0 \%)$      &$0.0\% (0.0 \%)$    &  $100.0\% (82.0\%)$      & $2.0\% (0.0 \%)$      & $0.0\% (0.0 \%)$     &  $98.0\% (81.6 \%)$       \\
\midrule
\multicolumn{6}{c}{Which document exhibits higher quality by considering the following aspects:} \\
\multicolumn{6}{c}{linguistic fluency, logical coherence, and information density?}\\
Human & LLM & Equal & Human & LLM & Equal  \\
$8.0 \% (0.0 \%)$      &$6.0\% (0.0 \%)$    &  $86.0\% (46.5 \%)$      & $4.0\% (0.0 \%)$      & $6.0\% (0.0 \%)$     &  $90.0\% (60. \%)$       \\
\bottomrule
\end{tabular}
}
\end{table}

\begin{table*}
\caption{Performance comparison of retrieval models for mixed human-written and Llama2-generated corpora on SciFact+AIGC and NQ320K+AIGC dataset. 
The \colorbox{RPD_pos1}{numbers} indicate that retrieval models rank human-written documents in higher positions than LLM-generated documents (i.e., $\text{Relative}~~\Delta > 0\%$). Conversely, the \colorbox{RPD_neg1}{numbers} mean retrieval models rank LLM-generated documents in higher positions than human-written documents (i.e., $\text{Relative}~~\Delta \leq 0\%$). The intensity of the color reflects the extent of the difference. In the subsequent tables, we will continue with this color scheme.}
\label{tab:main_llama}
\vspace{-0.2cm}
\centering
\resizebox{1.0\textwidth}{!}{
\begin{tabular}{ccccccccccccccc}
\hline\hline
Model     & \multirow{2}{*}{Model}       & \multirow{2}{*}{Target Corpus}                         & \multicolumn{6}{c}{SciFact+AIGC} &    \multicolumn{6}{c}{NQ320K+AIGC}      \\
\cmidrule(lr){4-9} \cmidrule(lr){10-15}
Type      &             &                                & NDCG@1    & NDCG@3 & NDCG@5 & MAP@1  & MAP@3  & MAP@5  & NDCG@1  & NDCG@3 & NDCG@5 & MAP@1   & MAP@3  & MAP@5   \\
\hline
\multirow[m]{6}{*}{Lexical}    & \multirow[m]{3}{*}{TF-IDF}     & Human-Written                  & 22.0      & 36.9   & 39.7   & 21.2  & 33.0  & 34.7  & 7.1    & 11.0   & 12.3   & 7.1   & 10.0  & 10.8   \\
           &            & LLM-Generated                  & 17.0      & 33.8   & 37.2   & 16.2  & 29.5  & 31.5  & 3.4    & 8.1    & 9.4    & 3.4   & 7.0   & 7.7    \\
           &            & $\text{Relative}~~\Delta$ & \cellcolor{RPD_pos2}25.6      & \cellcolor{RPD_pos1}8.8    & \cellcolor{RPD_pos1}6.5    & \cellcolor{RPD_pos2}26.7  & \cellcolor{RPD_pos2}11.2  & \cellcolor{RPD_pos1}9.7   & \cellcolor{RPD_pos2}70.5   & \cellcolor{RPD_pos2}30.4   & \cellcolor{RPD_pos2}26.7   & \cellcolor{RPD_pos2}70.5  & \cellcolor{RPD_pos2}35.3  & \cellcolor{RPD_pos2}33.5   \\
\cline{2-15}           
           & \multirow[m]{3}{*}{BM25}       & Human-Written                  & 26.7      & 40.3   & 44.4   & 25.7  & 36.7  & 39.1  & 7.2    & 11.6   & 12.9   & 7.2   & 10.6  & 11.3   \\
           &            & LLM-Generated                  & 21.0      & 38.8   & 41.5   & 19.6  & 34.3  & 35.9  & 6.1    & 10.9   & 11.9   & 6.1   & 9.7   & 10.3   \\
           &            & $\text{Relative}~~\Delta$ & \cellcolor{RPD_pos2}23.9      & \cellcolor{RPD_pos0}3.8    & \cellcolor{RPD_pos1}6.8    & \cellcolor{RPD_pos2}26.9  & \cellcolor{RPD_pos1}6.8   & \cellcolor{RPD_pos1}8.5   & \cellcolor{RPD_pos2}16.5   & \cellcolor{RPD_pos1}6.2    & \cellcolor{RPD_pos1}8.1    & \cellcolor{RPD_pos2}16.5  & \cellcolor{RPD_pos1}8.9   & \cellcolor{RPD_pos1}9.3    \\
\hline           
\multirow[m]{12}{*}{Neural}     & \multirow[m]{3}{*}{ANCE}       & Human-Written                  & 15.3      & 30.1   & 32.7   & 14.2  & 26.2  & 27.7  & 22.2   & 41.2   & 44.6   & 22.2  & 36.9  & 38.8   \\
           &            & LLM-Generated                  & 24.7      & 35.8   & 37.7   & 23.3  & 32.4  & 33.6  & 29.1   & 45.9   & 49.0   & 29.1  & 42.0  & 43.8   \\
           &            & $\text{Relative}~~\Delta$ & \cellcolor{RPD_neg2}-47.0     & \cellcolor{RPD_neg2}-17.3  & \cellcolor{RPD_neg2}-14.2  & \cellcolor{RPD_neg2}-48.5 & \cellcolor{RPD_neg2}-21.2 & \cellcolor{RPD_neg2}-19.2 & \cellcolor{RPD_neg2}-26.9  & \cellcolor{RPD_neg2}-10.8  & \cellcolor{RPD_neg1}-9.4   & \cellcolor{RPD_neg2}-26.9 & \cellcolor{RPD_neg2}-12.9 & \cellcolor{RPD_neg2}-12.1  \\
\cline{2-15}           
           & \multirow[m]{3}{*}{BERM}       & Human-Written                  & 16.3      & 30.2   & 31.8   & 15.7  & 26.5  & 27.5  & 18.6   & 37.5   & 40.7   & 18.6  & 33.1  & 34.9   \\
           &            & LLM-Generated                  & 23.7      & 34.1   & 36.4   & 21.7  & 30.8  & 32.2  & 31.6   & 47.0   & 50.0   & 31.6  & 43.5  & 45.1   \\
           &            & $\text{Relative}~~\Delta$ & \cellcolor{RPD_neg2}-37.0     & \cellcolor{RPD_neg2}-12.1  & \cellcolor{RPD_neg2}-13.5  & \cellcolor{RPD_neg2}-32.1 & \cellcolor{RPD_neg2}-15.0 & \cellcolor{RPD_neg2}-15.7 & \cellcolor{RPD_neg2}-51.8  & \cellcolor{RPD_neg2}-22.5  & \cellcolor{RPD_neg2}-20.5  & \cellcolor{RPD_neg2}-51.8 & \cellcolor{RPD_neg2}-27.2 & \cellcolor{RPD_neg2}-25.5  \\
\cline{2-15}           
           & \multirow[m]{3}{*}{TAS-B}      & Human-Written                  & 20.0      & 40.2   & 43.1   & 19.5  & 35.2  & 36.9  & 25.7   & 45.4   & 48.8   & 25.7  & 40.9  & 42.8   \\
           &            & LLM-Generated                  & 31.7      & 44.8   & 47.5   & 29.7  & 41.1  & 42.7  & 27.6   & 46.5   & 50.0   & 27.6  & 42.2  & 44.2   \\
           &            & $\text{Relative}~~\Delta$ & \cellcolor{RPD_neg2}-45.3     & \cellcolor{RPD_neg2}-10.8  & \cellcolor{RPD_neg1}-9.7   & \cellcolor{RPD_neg2}-41.5 & \cellcolor{RPD_neg2}-15.5 & \cellcolor{RPD_neg2}-14.6 & \cellcolor{RPD_neg1}-7.1   & \cellcolor{RPD_neg0}-2.4   & \cellcolor{RPD_neg0}-2.4   & \cellcolor{RPD_neg1}-7.1  & \cellcolor{RPD_neg0}-3.1  & \cellcolor{RPD_neg0}-3.2   \\
\cline{2-15}           
           & \multirow[m]{3}{*}{Contriever} & Human-Written                  & 24.0      & 43.7   & 47.8   & 23.3  & 38.8  & 41.2  & 25.9   & 48.5   & 51.9   & 25.9  & 43.3  & 45.3   \\
           &            & LLM-Generated                  & 31.0      & 47.8   & 50.5   & 29.6  & 43.2  & 44.8  & 32.5   & 51.9   & 55.4   & 32.5  & 47.5  & 49.4   \\
           &            & $\text{Relative}~~\Delta$ & \cellcolor{RPD_neg2}-25.5     & \cellcolor{RPD_neg1}-9.0   & \cellcolor{RPD_neg1}-5.5   & \cellcolor{RPD_neg2}-23.8 & \cellcolor{RPD_neg2}-10.7 & \cellcolor{RPD_neg1}-8.4  & \cellcolor{RPD_neg2}-22.6  & \cellcolor{RPD_neg1}-6.8   & \cellcolor{RPD_neg1}-6.5   & \cellcolor{RPD_neg2}-22.6 & \cellcolor{RPD_neg1}-9.3  & \cellcolor{RPD_neg1}-8.7  \\
\hline\hline               
              
\end{tabular}
}
\end{table*}

\subsubsection{Retrieval Performance Evaluation}
To further validate the accuracy of the relevancy label assignments, we conduct an evaluation of retrieval models on the human-written corpus and the LLM-generated corpus, respectively. The following representative retrieval models are adopted in the experiments: (1) Lexical Retrieval Models: \textbf{TF-IDF}~\cite{sparck1972statistical} and \textbf{BM25}~\cite{robertson2009probabilistic} and (2) Neural Retrieval Models: \textbf{ANCE}~\cite{xiong2020approximate}, \textbf{BERM}~\cite{xu-etal-2023-berm}, \textbf{TAS-B}~\cite{hofstatter2021efficiently}, \textbf{Contriever}~\cite{izacard2021unsupervised}. 

The results on each sole source corpus on the proposed two new benchmarks are presented in \autoref{tab:quality2_llama2_gaps}. It is evident that all retrieval models exhibit no significant performance discrepancies in terms of various ranking metrics between the human-written and LLM-generated corpora across all datasets. This observation reinforces the confidence in the quality of our newly constructed datasets.

\subsubsection{Human Evaluation}
Note that in our constructed datasets, LLMs were instructed to rewrite human-written texts based solely on the original human-written text, without any query-related input, thereby \textit{preventing the additional query-specific information during rewriting}.
Moreover, to further verify this, we conduct a human evaluation.
Specifically, we randomly select 50 <query, human-written document, LLM-generated document> triples from each dataset. The human annotators, comprising the authors and their highly educated colleagues, are asked to determine which document is more semantically relevant to the given query. The options are ``Human'', ``LLM'', or ``Equal''. During the evaluation, annotators are unaware of the source of each document. Each triple is labeled at least by three different annotators, with the majority vote determining the final label. 
The results in \autoref{tab: human_eval}, confirm that both sources of texts have almost the same semantic relevance to the given queries, which guarantees the fairness of our following exploration of source bias.

Additionally, we also conduct further human evaluations specifically focused on text quality. The human annotators are asked to determine ``Which document exhibits higher quality by considering the following aspects: linguistic fluency, logical coherence, and information density?''
The notation process is the same as above, and the results are summarized in \autoref{tab: human_eval}.
The results indicate no significant distinction between LLM-generated and human-written content on text quality, demonstrating consistency across both sources.
In fact, we also analyze the data cases and find that LLMs typically alter only parts of the vocabulary, leading to minor stylistic differences without impacting the core content, which can be further verified with these human evaluations.

\section{RQ2: Uncovering Source Bias}

In this section, we conduct extensive experiments on the constructed datasets to explore the source bias from various aspects. With the constructed simulated environment, we first introduce the evaluation metrics to quantify the severity of source bias. We then conduct experiments with different retrieval models on both the first-stage retrieval and the second-stage re-ranking.

\subsection{Evaluation Metrics for Source Bias}

To quantitatively explore source bias, we calculate ranking metrics,  targeting separately either human-written or LLM-generated corpus.
Specifically, for each query, an IR model produces a ranking list that comprises documents from mixed corpora. 
We then calculate top-$K$ Normalized Discounted Cumulative Gain (NDCG@$K$) and Mean Average Precision (MAP@$K$), for $K \in \{1,3,5\}$, independently for each corpus source. 
When assessing one corpus (e.g., human-written), documents from the other (e.g., LLM-generated) are treated as non-relevant, though the original mixed-source ranking order is maintained. 
This approach allows us to independently assess the performance of IR models on each corpus source. 

To better normalize the difference among different benchmarks, we also introduce the relative percentage difference as follows:
$$
 \text{Relative $\Delta$} = \frac{\texttt{Metric}_\text{Human-written} - \texttt{Metric}_\text{LLM-generated}}{\frac{1}{2}(\texttt{Metric}_\text{Human-written} +\texttt{Metric}_\text{LLM-generated})} \times 100\%,
$$
where the $\texttt{Metric}$ can be NDCG@$K$ and MAP@$K$. Note that Relative $\Delta > 0$ means retrieval models rank human-written texts higher, and  Relative $\Delta < 0$ indicates LLM-generated texts are ranked higher. The greater the absolute value of Relative $\Delta$, the greater the ranking performance difference between two sourced content.

\subsection{Bias in Neural Retrieval Models} \label{sec:results_chatgpt}

\begin{table}[t]
\caption{Performance comparison of different neural retrieval models for mixed human-written and ChatGPT-generated corpora on SciFact+AIGC dataset. }
\label{tab:gpt_scifact}
\centering
\resizebox{1\linewidth}{!}{
\begin{tabular}{cccccccc}
\hline\hline
Model      & Target Corpus                         &  NDCG@1    & NDCG@3 & NDCG@5 & MAP@1  & MAP@3  & MAP@5   \\
\hline
\multirow[m]{3}{*}{TF-IDF} & Human-Written                  & 22.7 & 36.5 & 39.5 & 22.0 & 32.8 & 34.6 \\
       & LLM-Generated                  & 16.7 & 34.9 & 37.1 & 16.0 & 30.2 & 31.4 \\
       & $\text{Relative}~~\Delta$ & \cellcolor{RPD_pos2}30.5 & \cellcolor{RPD_pos0}4.5  & \cellcolor{RPD_pos1}6.3  & \cellcolor{RPD_pos2}31.6 & \cellcolor{RPD_pos1}8.3  & \cellcolor{RPD_pos1}9.7  \\
\hline       
\multirow[m]{3}{*}{BM25}   & Human-Written                  & 24.3 & 38.5 & 42.7 & 23.7 & 34.8 & 37.3 \\
       & LLM-Generated                  & 24.3 & 40.2 & 42.7 & 23.1 & 35.8 & 37.3 \\
       & $\text{Relative}~~\Delta$ & \cellcolor{RPD_neg0}0.0  & \cellcolor{RPD_neg0}-4.3 & \cellcolor{RPD_neg0}0.0  & \cellcolor{RPD_pos0}2.6  & \cellcolor{RPD_neg0}-2.8 & \cellcolor{RPD_neg0}0.0  \\
\hline         
 \multirow[m]{3}{*}{ANCE}       & Human-Written                  & 18.0     & 30.8  & 33.8  & 16.5  & 27.2  & 29.0   \\
           & LLM-Generated                  & 24.7     & 35.6  & 37.4  & 24.0  & 32.7  & 33.7   \\
           & $\text{Relative}~~\Delta$ & \cellcolor{RPD_neg2}-31.4    & \cellcolor{RPD_neg2}-14.5 & \cellcolor{RPD_neg2}-10.1 & \cellcolor{RPD_neg2}-37.0 & \cellcolor{RPD_neg2}-18.4 & \cellcolor{RPD_neg2}-15.0  \\
\hline           
 \multirow[m]{3}{*}{BERM}       & Human-Written                  & 16.3     & 29.9  & 32.3  & 14.8  & 26.0  & 27.4   \\
                         & LLM-Generated                  & 22.7     & 32.5  & 35.3  & 21.9  & 29.7  & 31.4   \\
                    & $\text{Relative}~~\Delta$ & \cellcolor{RPD_neg2}-32.8    & \cellcolor{RPD_neg1}-8.3  & \cellcolor{RPD_neg1}-8.9  & \cellcolor{RPD_neg2}-38.7 & \cellcolor{RPD_neg2}-13.3 & \cellcolor{RPD_neg2}-13.6   \\
\hline               
 \multirow[m]{3}{*}{TAS-B}      & Human-Written                  & 23.0     & 41.5  & 44.4  & 22.2  & 36.9  & 38.6   \\
           & LLM-Generated                  & 28.7     & 45.5  & 46.7  & 27.2  & 40.9  & 41.6   \\
           & $\text{Relative}~~\Delta$ & \cellcolor{RPD_neg2}-22.1    & \cellcolor{RPD_neg1}-9.2  & \cellcolor{RPD_neg1}-5.0  & \cellcolor{RPD_neg2}-20.2 & \cellcolor{RPD_neg2}-10.3 & \cellcolor{RPD_neg1}-7.5   \\
\hline           
 \multirow[m]{3}{*}{Contriever} & Human-Written                  & 24.0     & 44.0  & 47.2  & 23.3  & 39.1  & 41.0   \\
           & LLM-Generated                  & 33.0     & 48.3  & 50.6  & 31.3  & 44.0  & 45.4   \\
           & $\text{Relative}~~\Delta$ & \cellcolor{RPD_neg2}-31.6    & \cellcolor{RPD_neg1}-9.3  & \cellcolor{RPD_neg1}-7.0  & \cellcolor{RPD_neg2}-29.3 & \cellcolor{RPD_neg2}-11.8 & \cellcolor{RPD_neg2}-10.2  \\
\hline\hline           
\end{tabular}
}
\end{table}

In our assessment of various retrieval models on SciFact+AIGC and NQ320K+AIGC datasets, we observe distinct phenomena when evaluating against human-written and LLM-generated corpora, as reported in \autoref{tab:main_llama}. Our key findings are as follows:

\textbf{Lexical models prefer human-written texts.} Lexical models like TF-IDF and BM25 show a tendency to favor human-written texts over LLM-generated texts across most ranking metrics in both datasets.
A plausible explanation for this phenomenon lies in the term-based distinctions between text generated by LLMs and human-written content, as evident in~\autoref{fig:analysis_of_jaccard_overlap}. Additionally, the queries are crafted by humans and thus exhibit a style more closely aligned with human-written text.

\textbf{Neural retrievers are biased towards LLM-generated texts.} Neural models, which rely on semantic matching with PLMs, demonstrate a pronounced preference for LLM-generated texts, often performing over 30\% better on these compared to human-written texts. 
These findings suggest an inherent bias in neural retrievers towards LLM-generated text, which we named the \textbf{source bias}.
This source bias may stem from PLMs-based neural retrievers and LLMs sharing similar Transformer-based architectures~\cite{vaswani2017attention} and pretraining approaches, leading to potential exploitation of \textit{semantic shortcuts} in LLM-generated text during semantic matching. Additionally, LLMs seem to semantically compress information in a manner that makes it more comprehensible to neural models. A deeper exploration into the causes of source bias is presented in the following section.

To strengthen our conclusion that \textbf{source bias is not limited to any specific LLM}, we extend our investigation to include ChatGPT, another widely adopted and nearly state-of-the-art LLM. We employ ChatGPT to generate a corpus using the same prompts as those utilized with Llama2 in the above experiments. 
Subsequently, in~\autoref{tab:gpt_scifact}, we report the evaluation results on the SciFact+AIGC dataset, which contains both human-written and ChatGPT-generated texts. Once again, the results clearly indicate a bias within neural retrieval models, favoring the corpus generated by ChatGPT across all ranking metrics. This observation provides additional substantiation of the presence of source bias within these neural retrieval models.

Furthermore, we also explore the popular InstructGPT-prompts GitHub Repository, which includes several common prompts for rephrasing passages~\footnote{\url{https://github.com/kevinamiri/Instructgpt-prompts}}. The experimental results in Appendix~\ref{app:more_prompts} show varying degrees of source bias, indicating that common prompts can easily trigger source bias with LLM-generated content.
These findings highlight the notable presence of source bias in neural retrieval models towards LLM-generated content.

\begin{table}[t]
\caption{Bias evaluation of re-ranking models on SciFact+AIGC dataset. The re-ranking methods rerank the top-100 retrieved hits from a first-stage BM25 model. }
\label{tab:rerank_scifact}
\centering
\resizebox{1\linewidth}{!}{
\begin{tabular}{ccccc|ccc}
\hline\hline
\multirow[m]{2}{*}{Metrics} & \multirow[m]{2}{*}{Target Corpus} & \multicolumn{3}{c|}{Llama2-generated} &  \multicolumn{3}{c}{ChatGPT-generated} \\
\cmidrule(lr){3-5} \cmidrule(lr){6-8}
 &                          & BM25  & +MiniLM & +monoT5  & BM25  & +MiniLM & +monoT5\\
\hline
\multirow[m]{3}{*}{NDCG@1}  & Human-Written                  & 26.7 & 21.3   & 19.7 & 24.3                & 18.3                & 21.3       \\
        & LLM-Generated                  & 21.0 & 32.7   & 39.7   & 24.3                & 35.7                & 39.3     \\
        & $\text{Relative}~~\Delta$ & \cellcolor{RPD_pos2}23.9 & \cellcolor{RPD_neg2}-42.2  & \cellcolor{RPD_neg2}-67.3    & \cellcolor{RPD_neg0}0.0                 & \cellcolor{RPD_neg2}-64.4               & \cellcolor{RPD_neg2}-59.4   \\
\hline        
\multirow[m]{3}{*}{NDCG@3}  & Human-Written                  & 40.3 & 42.8   & 45.9   & 38.5                & 41.4                & 46.4        \\
        & LLM-Generated                  & 38.8 & 47.8   & 52.9    & 40.2                & 50.1                & 54.2      \\
        & $\text{Relative}~~\Delta$ & \cellcolor{RPD_pos0}3.8  & \cellcolor{RPD_neg2}-11.0  & \cellcolor{RPD_neg2}-14.2   & \cellcolor{RPD_neg0}-4.3                & \cellcolor{RPD_neg2}-19.0               & \cellcolor{RPD_neg2}-15.5    \\
\hline        
\multirow[m]{3}{*}{NDCG@5}  & Human-Written                  & 44.4 & 46.9   & 49.0    & 42.7                & 45.6                & 48.9         \\
        & LLM-Generated                  & 41.5 & 50.2   & 54.7   & 42.7                & 53.0                & 56.1     \\
        & $\text{Relative}~~\Delta$ & \cellcolor{RPD_pos1}6.8  & \cellcolor{RPD_neg1}-6.8   & \cellcolor{RPD_neg2}-11.0    & \cellcolor{RPD_neg0}0.0                 & \cellcolor{RPD_neg2}-15.0               & \cellcolor{RPD_neg2}-13.7   \\
\hline        
\multirow[m]{3}{*}{MAP@1}   & Human-Written                  & 25.7 & 20.8   & 18.9    & 23.7                & 17.9                & 20.5      \\
        & LLM-Generated                  & 19.6 & 30.8   & 37.8   & 23.1                & 33.8                & 37.8     \\
        & $\text{Relative}~~\Delta$ & \cellcolor{RPD_pos2}26.9 & \cellcolor{RPD_neg2}-38.8  & \cellcolor{RPD_neg2}-66.7  & \cellcolor{RPD_pos0}2.6                 & \cellcolor{RPD_neg2}-61.5               & \cellcolor{RPD_neg2}-59.3     \\
\hline        
\multirow[m]{3}{*}{MAP@3}   & Human-Written                  & 36.7 & 37.5   & 39.7  & 34.8                & 35.8                & 40.3      \\
        & LLM-Generated                  & 34.3 & 43.6   & 48.9    & 35.8                & 45.9                & 50.0     \\
        & $\text{Relative}~~\Delta$ & \cellcolor{RPD_pos1}6.8  & \cellcolor{RPD_neg2}-15.0  & \cellcolor{RPD_neg2}-20.8   & \cellcolor{RPD_neg0}-2.8                & \cellcolor{RPD_neg2}-24.7               & \cellcolor{RPD_neg2}-21.5     \\
\hline        
\multirow[m]{3}{*}{MAP@5}   & Human-Written                  & 39.1 & 40.0   & 41.6      & 37.3                & 38.3                & 41.7    \\
        & LLM-Generated                  & 35.9 & 45.0   & 50.1    & 37.3                & 47.6                & 51.4      \\
        & $\text{Relative}~~\Delta$ & \cellcolor{RPD_pos1}8.5  & \cellcolor{RPD_neg2}-11.8  & \cellcolor{RPD_neg2}-18.5  & \cellcolor{RPD_neg0}0.0                 & \cellcolor{RPD_neg2}-21.7               & \cellcolor{RPD_neg2}-20.8     \\
\hline\hline        
\end{tabular}
}
\end{table}

\subsection{Bias in Re-Ranking Stage}

In a typical IR system, there are two primary stages of document filtering. The first stage involves a retriever, responsible for document recall, while the second stage employs a re-ranker, which fine-tunes the ordering of documents within the initially retrieved set. 
While we have revealed the presence of the source bias in the first stage, a natural pivotal research question remains: does this bias also manifest in the re-ranking stage?  To delve into this, we select two representative and state-or-the-art re-ranking models: \textbf{MiniLM}~\cite{wang2020minilm} and \textbf{monoT5}~\cite{nogueira2020document} to rerank the top-100 document list retrieved by a first-stage BM25 model. 

The results on the SciFact+AIGC dataset with Llama-generated corpus and ChatGPT-generated corpus are presented in \autoref{tab:rerank_scifact}. From the results, while even the first-stage retrievers (BM25) may exhibit a preference for human-written content, the second-stage re-rankers once again demonstrate a bias in favor of LLM-generated content. Remarkably, the bias in re-ranking models appears to be more severe, as evidenced by the relative percentage difference of $-67.3\%$ and $-59.4\%$ in NDCG$@1$ for monoT5, respectively. These findings further confirm the pervasiveness of source bias in neural ranking models that rely on PLMs, regardless of the first retrieval stage or second re-ranking stage.

\section{RQ3: The Cause of Source Bias}
In this section, we delve deeper into why neural retrieval models exhibit source bias.
Our objective is to determine whether the LLM-generated texts, characterized by reduced noise and more concentrated semantic topics, are inherently easier for neural retrieval models to semantically match.
We conduct a series of analyses from the perspective of text compression and provide valuable insights.

\begin{figure}[t]  
    \centering    
    \includegraphics[width=0.95\linewidth]{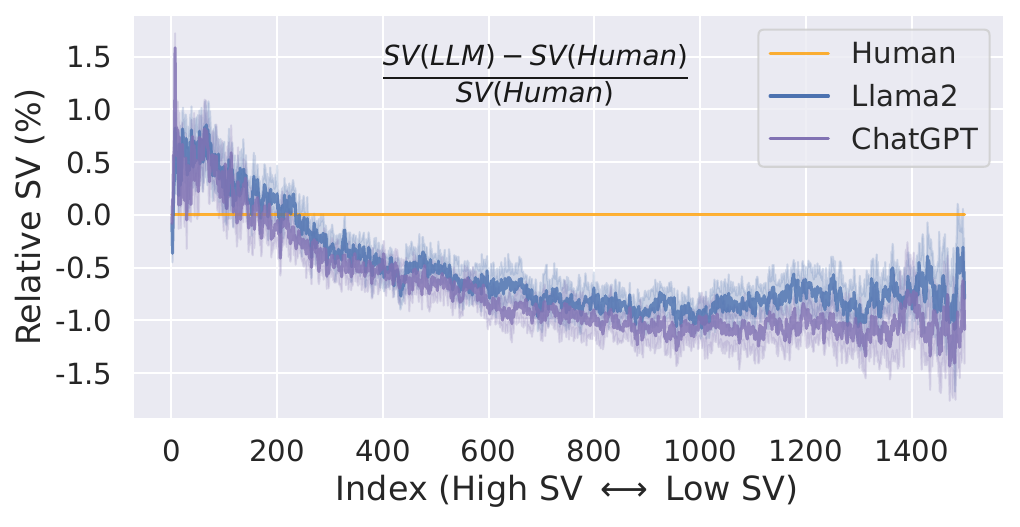}
    \caption{Comparision of the relative singular value (SV) of the different corpus after SVD. The singular values are sorted in descending order from left to right.}
    \label{fig:scifact_svd}  
\end{figure}

\subsection{Viewpoint from Text Compression}
We first explore the cause of source bias from a compression perspective, drawing inspiration from recent studies that suggest LLMs are lossless compressors~\cite{deletang2023language}. We hypothesize that LLMs efficiently focus on essential information, minimizing noise during generation, in contrast to human-written texts, which may include more diverse topics and incidental noise. To verify this, we employ Singular Value Decomposition (SVD)~\cite{klema1980singular} to compare topic concentration and noise in human-written and LLM-generated texts. The dimension of the SVD corresponds to the maximum number of topics, and the singular value associated with each topic represents its strength. High singular values predominantly capture primary topic information, whereas low singular values indicate noise.

Specifically, we utilize the OpenAI embedding model to obtain embedding matrices for each corpus in the SciFact+AIGC dataset and then conduct SVD. The resulting singular values are arranged in descending order, and their comparison to the human-written corpus is visualized in \autoref{fig:scifact_svd}. As we can see, LLM-generated texts exhibit larger singular values at the top large singular values, while smaller singular values at the tail small singular values. 
This observation suggests that LLM-generated texts tend to have more focused semantics with less noise, rendering them more suitable for precise semantic matching. 
In contrast, human-written texts often contain a wider range of latent topics and higher levels of noise, making them harder for neural retrievers to understand. As a result, this difference in semantic concentration may contribute to the observed source bias in neural retrievers.

\begin{figure}[t]
    \includegraphics[width=0.8 \linewidth]{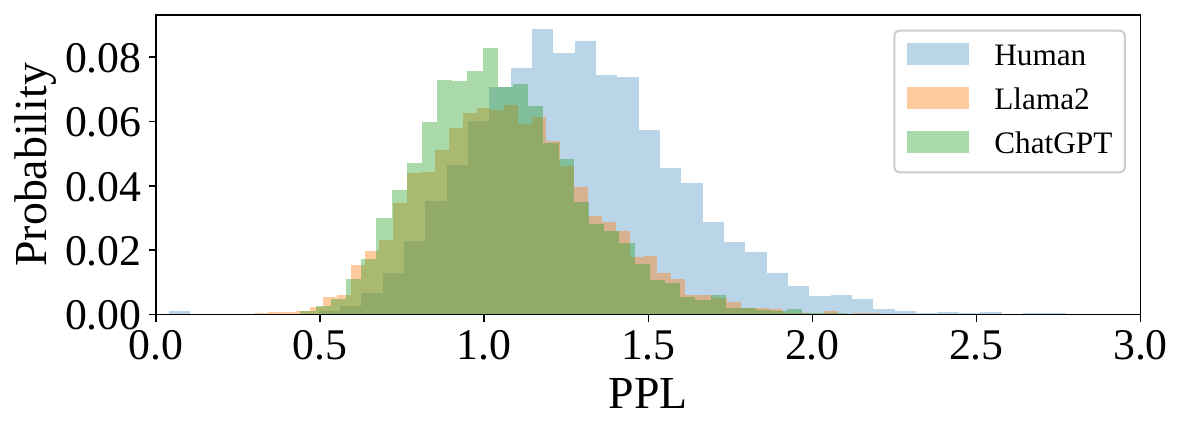}
    \label{fig:scifact_ppl_model}
    \vspace{-0.3cm}
    \caption{Comparision of the PPL of the different corpus.}
    \label{fig:scifact_ppl}
\end{figure}

\subsection{Further Analysis from Perplexity}

Considering that most modern neural retrievers are grounded on PLMs~\cite{yates2021pretrained, guo2020deep, zhao2022dense}, such as BERT~\cite{devlin-etal-2019-bert}, Roberta~\cite{liu2019roberta}, and T5~\cite{raffel2020exploring}, we analyze the perplexity of PLMs to further support the conclusion above from the viewpoint of compression that LLM-generated texts can be better understood by PLMs.
Perplexity is an important metric for evaluating how well a language model can understand a given text~\cite{azzopardi2003investigating, wang2019language}. For a specific language model (LM) and a document $d=(d_0,d_1, \cdots, d_S)$, the log perplexity is defined as the exponentiated average negative log-likelihood of each token in the tokenized sequence of $d$\footnote{For simplicity, we denote the log perplexity as PPL.}:
$$
 \text{PPL} (d) = -\frac{1}{S} \left(\sum_{s=1}^{S} \log P_\text{LM}(d_s|\text{context})\right),
$$
where $S$ is the token length of text $d$ and $P_\text{LM}(d_s)$ is the predicted likelihood of the $s$-th token conditioned on the context.
Lower perplexity suggests more confidence and understanding of LM for text patterns, while higher perplexity implies greater uncertainty in predictions, often arising from complex or unpredictable text patterns.

Using the most widely-used LM, BERT~\cite{devlin-etal-2019-bert}, as an example, we employ it to calculate the PPL for different corpus. As BERT is not an autoregressive LM, we follow standard practices~\cite{wang2021reinforcing, wang2019bert} to calculate the likelihood of each token conditioned on the other tokens, i.e., 
$$
P_\text{LM}(d_s|\text{context}) :=  P_\text{BERT}(d_s|d_{\leq S \backslash \{s\}}).
$$
The distribution of perplexity for different corpus in the SciFact+AIGC dataset is shown in \autoref{fig:scifact_ppl}. Notably, LLM-generated texts consistently exhibit significantly lower perplexity, indicating enhanced comprehensibility and higher confidence from BERT's perspective. 
Consequently, PLMs-based neural retrievers can more effectively model the semantics of LLM-generated texts, leading to the observed source bias in favor of LLM-generated texts.

In Appendix~\ref{app: theoretical_ana}, we also provide a theoretical analysis to illustrate and verify the observation in \autoref{fig:scifact_ppl} that LLM-generated texts have a smaller perplexity than human-written texts.

\section{RQ4: Mitigating Source Bias}
In this section, we propose a simple but effective approach to mitigate source bias by introducing a debiased constraint to the optimization objective. In this way, we can force the neural IR models to focus on modeling semantic relevance rather than the inherent semantic shortcut of the LLM-generated content.

\begin{figure}[t]
  \subfigure[Results on mixed human-written and Llama2-generated corpora]
    {
    \includegraphics[width=1\columnwidth]{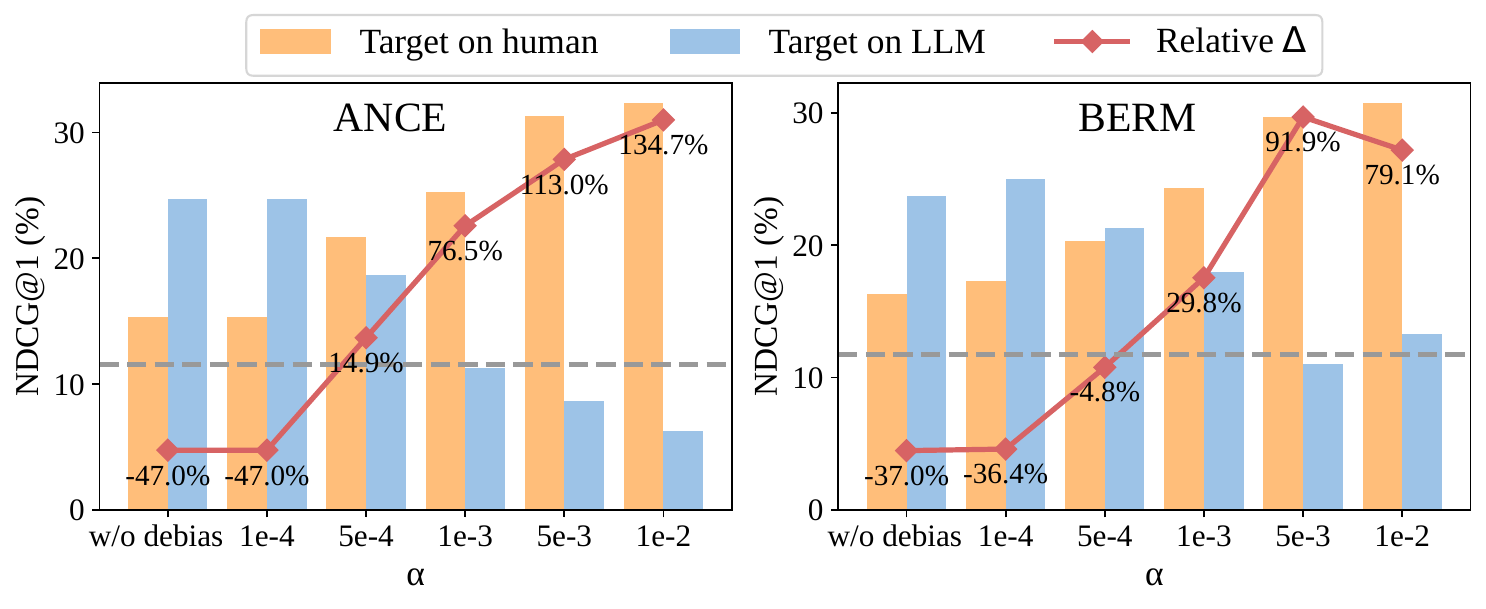}
    \label{fig:debias_human_llama2_mixed_ndcg1}
   }
  \subfigure[Results on mixed human-written and ChatGPT-generated corpora]{
    \includegraphics[width=1\columnwidth]{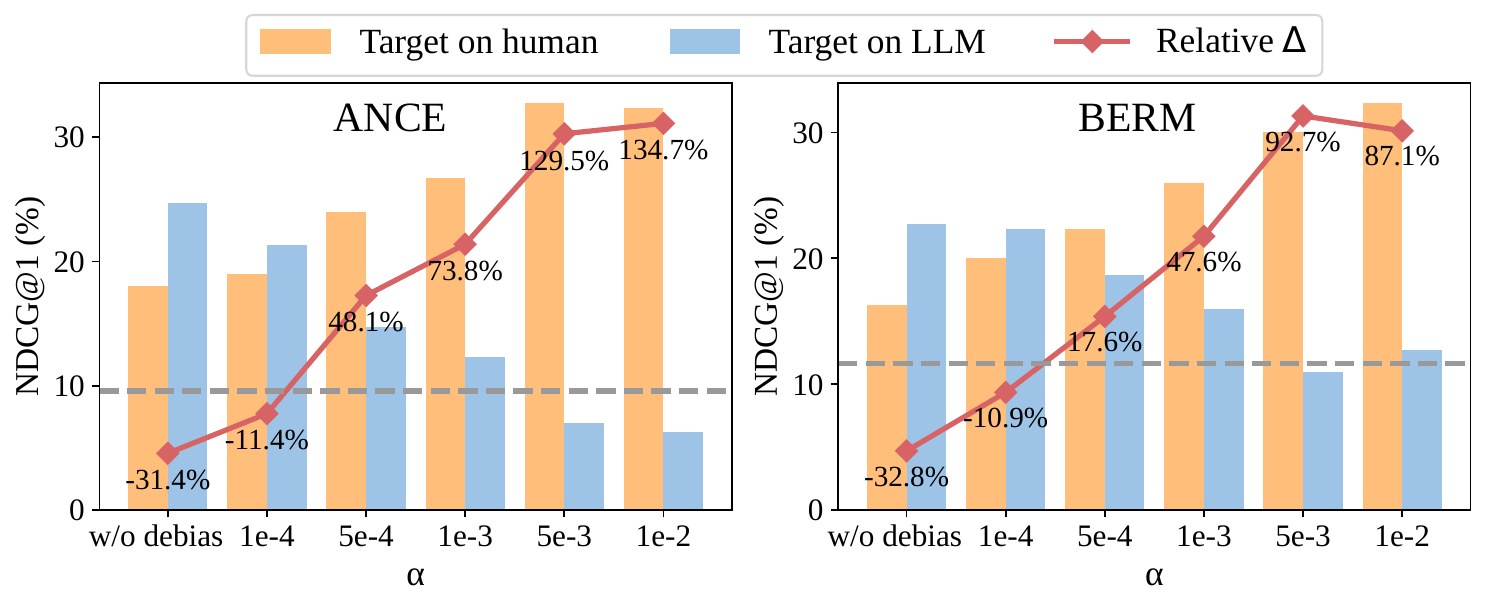}
    \label{fig:debias_human_chatgpt_mixed_ndcg1}
  }
    \caption{Performance comparison (NDCG@$1$) of neural models on SciFact+AIGC dataset with different debiased co-efficient $\alpha$. 
    The grey dashed line represents Relative $\Delta=0$.}
    \label{fig:debias_mixed}
\end{figure}

\subsection{Our Method: A Debiased Constraint}
Our earlier findings of source bias indicate that neural retrievers tend to rank LLM-generated documents in higher positions.
Thus, the motivation of our debiased method is straightforward, which is to force the retrieval models to focus on modeling the semantic relevance and not assign higher predicted relevance scores to the LLM-generated documents.
Specifically, following the practice in Section~\ref{sec: data_gen}, we first generate the corresponding LLM-generated corpus $\mathcal{C}^G$ for the original human-written training corpus $\mathcal{C}^H$. In this way, we can get the new paired training data $\mathcal{D} = \{(q_m, d^H_m, d^G_m)\}_{m=1}^{M}$, where each element $(q_m, d^H_m, d^G_m)$ is a <query, human-written document, LLM-generated document> triplet. $d^H_m$ and $d^G_m$ are the corresponding human-written and LLM-generated relevant documents for the query $q$, respectively. Then we introduce the debiased constraint, which can be defined as
\begin{equation} \label{loss:debiased_loss}
    \mathcal{L}_\text{debias} = \sum_{(q_m, d^H_m, d^G_m) \in \mathcal{D}} \max \{0, \hat{r}(q,d^G; \Theta) - \hat{r}(q,d^H; \Theta)\}
\end{equation}
where $\hat{r}(q,d^G; \Theta)$ and $\hat{r}(q,d^H; \Theta)$ are the predicted relevance scores of $(q,d^G)$ and $(q,d^H)$ by the retrieval models with parameters $\Theta$, respectively.
This constraint can penalize biased samples when the predicted relevance score of $(q, d^G)$ is greater than that of $(q, d^H)$. 

Based on the debiased constraint  defined in \eqref{loss:debiased_loss}, we can define the final loss for training an unbiased neural retriever:
\begin{equation}
    \mathcal{L} = \mathcal{L}_\text{rank} + \alpha \mathcal{L}_\text{debias}
\end{equation}
where the $\mathcal{L}_\text{rank}$ can be any common-used loss for the ranking task, e.g., contrastive loss or regression loss~\cite{zhao2022dense, guo2020deep, guo2022semantic}. And $\alpha$ is the debiased co-efficient that can balance the ranking performance and the degree of the source bias. The larger $\alpha$ indicates the greater penalty on the biased samples, leading to the retriever being more likely to rank the human-written texts in higher positions.

\subsection{Results and Analysis}
To evaluate the effectiveness of our proposed debiased method, we equip the debiased constraint defined in Eq.~\eqref{loss:debiased_loss} to two representative neural retrievers: ANCE~\cite{xiong2020approximate} and BERM~\cite{xu-etal-2023-berm}. In the experiments, we vary the debiased co-efficient $\alpha$ within the range of $\{1e\text{-}4, 5e\text{-}4, 1e\text{-}3, 5e\text{-}3, 1e\text{-}2\}$. The original retrieval models learned without the debiased constraint are denoted as ``w/o debias''. The results on the SciFact+AIGC dataset are presented in \autoref{fig:debias_mixed}.

\begin{figure}[t]
  \subfigure[ANCE]
    {
    \includegraphics[width=0.48\columnwidth]{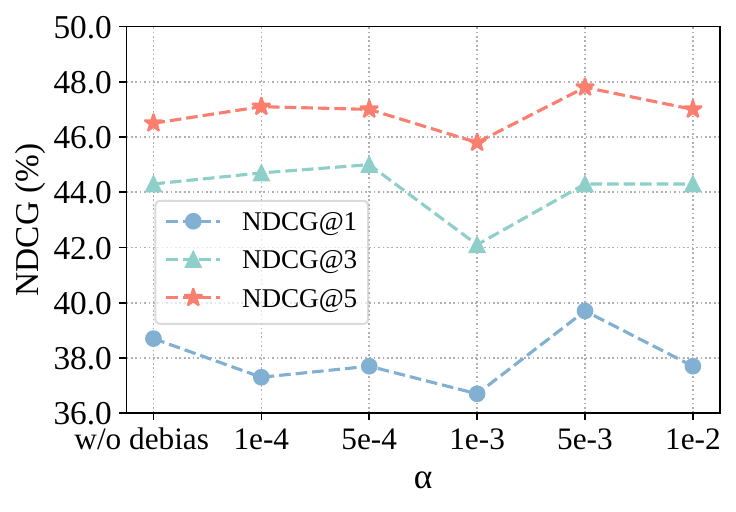}
    \label{fig:ance_only_human_ndcg}
   }
  \hspace{-0.1in}
  \subfigure[BERM]{
    \includegraphics[width=0.48\columnwidth]{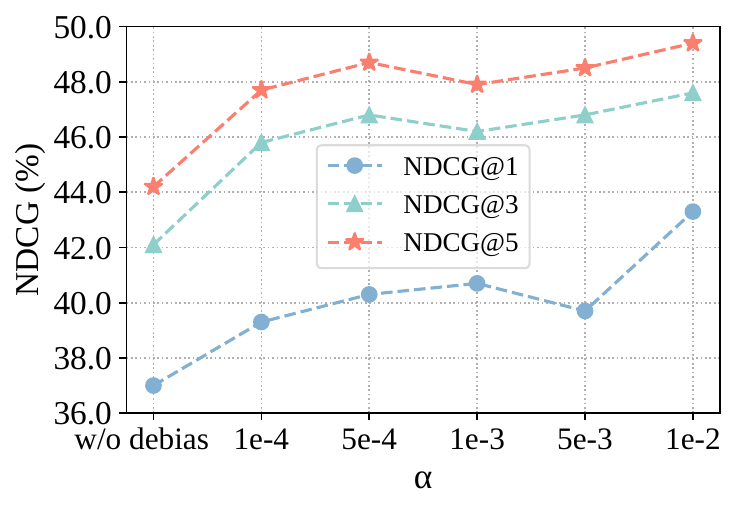}
    \label{fig:berm_only_human_ndcg}
  }
    \caption{Performance comparison of neural retrievers on only human-written SciFact dataset with different co-efficient $\alpha$ in our proposed debiased method.}
    \label{fig:debias_only_human}
\end{figure}

As we can see, as the debiased co-efficient $\alpha$ increases, the Relative $\Delta$ gradually shifts from negative to positive across almost all metrics and mixed datasets. This trend indicates that the neural retrieval models can rank human-written text higher than LLM-generated text with large $\alpha$. This can be attributed to the inclusion of our debiased constraint into the learning objective, which can penalize the biased samples and compel the retrieval models not to assign higher predicted relevance scores to LLM-generated content. 
Moreover, as shown in \autoref{fig:debias_only_human}, our method not only maintains the retrieval performance on the sole human-written corpus but also provides improvements, especially with BERM as the backbone.
This improvement is likely due to the inclusion of LLM-generated samples, which might enhance the model's ability to discern relevance among similar documents.

In summary, these empirical results have demonstrated the efficacy of our proposed debiased method in mitigating source bias to different extents by adjusting the debiased coefficient $\alpha$. 
This flexibility allows for customizing debiasing mechanisms to meet diverse perspectives and demands.
Notably, the decision to maintain equality between the two content sources or favor human-written content can be tailored based on specific requirements and environmental considerations. For example, users may not mind the content's source if it is of high quality and fulfills their informational needs. However, bias becomes a significant issue when we aim to credit content providers and encourage more creation, impacting the sustainability of the content creation ecosystem. 
The optimal strategy for enhancing the sustainable development of the IR ecosystem remains an open question for further exploration.

\section{Discussion: Sounding the Alarm}

Through a rigorous series of experiments and thorough analysis, we have identified that neural retrieval models demonstrate clear preferences for LLM-generated texts, referred to as source bias. 
This bias, with the burgeoning proliferation of LLMs and AIGC, may raise significant concerns for a variety of aspects.

\textbf{First}, the presence of source bias poses a significant risk of gradually rendering human-written content less accessible, potentially causing a disruption in the content ecosystem. More severely, the concern is escalating with the growing prevalence of LLM-generated content online\cite{hanley2023machine, bengio2023managing}.
\textbf{Second}, there is the risk that source bias may amplify the spread of misinformation, especially considering the potential of LLMs to generate deceptive content, whether intentionally or not~\cite{chen2023can, Su2023FakeND, pan2023risk, aslett2023online}.
\textbf{Third}, source bias may be maliciously exploited to attack against neural retrieval models within today's search engines, creating a precarious vulnerability that could be weaponized by malicious actors, reminiscent of earlier web spam link attacks against PageRank~\cite{gyongyi2004combating}.

As discussed above, since LLMs can be readily instructed to generate texts at scale, source bias presents potential tangible and serious threats to the ecosystem of web content, public trust, and online safety. We hope this discussion will sound the alarm regarding the risks posed by source bias in the LLM era.

\section{Related Work}

\paratitle{Large Language Models for IR.}
The emergence of large language models (LLMs)~\cite{zhao2023survey, wei2022emergent, yang2023harnessing} has ushered in a transformative era across various research domains, such as natural language processing (NLP)~\cite{bubeck2023sparks, bang2023multitask}, education~\cite{guo2023close, Nunes2023EvaluatingGA}, recommender systems~\cite{dai2023uncovering, fan2023recommender}, finance~\cite{wu2023bloomberggpt, huang2023finbert}, and medicine~\cite{alberts2023large, thirunavukarasu2023large}. 
In the field of IR, much effort has also been made to utilize the remarkable knowledge and capabilities of LLMs to enhance IR systems~\cite{ai2023information, zhu2023large}. 
In the industry community, an exemplary successful application is New Bing\footnote{\url{https://www.bing.com/new}}, which is an LLM-powered search assistant that adeptly extracts information from various web pages and delivers concise summarized responses to user queries, thereby improving the search experience. In the research community, there has been a proactive exploration of integrating LLMs into the IR components, including query rewriters~\cite{srinivasan2022quill, wang2023query2doc}, retrievers~\cite{xu2022match, dai2022promptagator}, re-rankers~\cite{sun2023chatgpt, cho2023discrete}, and readers~\cite{shi2023replug, izacard2022few}. For a more comprehensive overview of the recent advancements in LLMs for IR, please refer to the recent survey~\cite{zhu2023large}.

\paratitle{Artificial Intelligence Generated Content.}
Artificial Intelligence Generated Content (AIGC) is a rapidly advancing field that involves the creation of content using advanced Generative AI (GAI)~\cite{cao2023comprehensive, wu2023ai, agnese2020survey}.
Unlike traditional content crafted by humans, AIGC can be generated at scale and in considerably less time~\cite{hanley2023machine, spitale2023ai}.
Recently, the development of LLMs and other GAI models has greatly improved the quality of AIGC content than before. 
For instance, LLMs such as ChatGPT have shown impressive abilities in generating human-like content~\cite{cao2023comprehensive, wu2023ai}. 
The DALL-E-3~\cite{betker2023improving}, another state-of-the-art text-to-image generation system, can follow user instructions to produce high-quality images.
Nevertheless, as AIGC becomes more prevalent across myriad domains, ethical concerns, and potential risks come into sharper focus~\cite{thakur-etal-2023-language, wang2023security}.
In fact, inevitably, the GAI models may generate content with bias and discrimination as the large training data always contain bias and toxicity~\cite{zhuo2023exploring, deshpande2023toxicity, bengio2023managing}. Furthermore, researchers have found that LLMs can be manipulated into generating increasingly deceptive misinformation, posing challenges to online safety~\cite{jiang2023disinformation, Su2023FakeND, chen2023can}.
In addition, some recent studies indicate that training GAI models with synthetic data could result in the collapse of the next-generation models~\cite{alemohammad2023self, shumailov2023model, briesch2023large}.
Thus, AIGC is a double-edged sword that requires cautious handling.

\section{Conclusion and Future Work}

In this paper, we provide a preliminary analysis of the impact of the proliferation of generated content on IR systems, which is a pressing and emerging problem in the LLM era.
We first introduce two new benchmarks, SciFact+AIGC and NQ320K+AIGC, and build an environment for evaluating IR models in scenarios where the corpus comprises both human-written and LLM-generated texts. Through extensive experiments within this environment, we uncover an unexpected bias of neural retrieval models favoring LLM-generated text. Moreover, we provide an in-depth analysis of this bias from the perspective of text compression.
We also introduce a plug-and-play debiased strategy, which shows the potential to mitigate the source bias to different degrees.
Finally, we discuss the crucial concerns and potential risks of this bias to the whole web ecosystem.

Our study offers valuable insights into several promising directions for future research, including exploring source bias in other information systems (e.g., recommender systems and advertising systems) and examining source bias in neural models towards AIGC data across multiple data modalities, not limited to text. Moreover, uncovering the root cause of the source bias and thus further mitigating it are difficult but crucial research directions.

\begin{acks}
This work was funded by the National Key R\&D Program of China (2023YFA1008704), the National Natural Science Foundation of China (No. 62377044, 62276248, 62376275, 62076234), Beijing Natural Science Foundation (No. 4222029), Beijing Key Laboratory of Big Data Management and Analysis Methods, Major Innovation \& Planning Interdisciplinary Platform for the  ``Double-First Class'' Initiative, PCC@RUC, funds for building world-class universities (disciplines) of Renmin University of China, and the Youth Innovation Promotion Association CAS under Grants No.2023111.
This work was supported by the Fundamental Research Funds for the Central Universities, and the Research Funds of Renmin University of China (RUC24QSDL013).
We thank all the anonymous reviewers for their positive and insightful comments.
\end{acks}

\clearpage

\bibliographystyle{ACM-Reference-Format}
\balance
\bibliography{ref}


\begin{thebibliography}{76}


\ifx \showCODEN    \undefined \def \showCODEN     #1{\unskip}     \fi
\ifx \showDOI      \undefined \def \showDOI       #1{#1}\fi
\ifx \showISBNx    \undefined \def \showISBNx     #1{\unskip}     \fi
\ifx \showISBNxiii \undefined \def \showISBNxiii  #1{\unskip}     \fi
\ifx \showISSN     \undefined \def \showISSN      #1{\unskip}     \fi
\ifx \showLCCN     \undefined \def \showLCCN      #1{\unskip}     \fi
\ifx \shownote     \undefined \def \shownote      #1{#1}          \fi
\ifx \showarticletitle \undefined \def \showarticletitle #1{#1}   \fi
\ifx \showURL      \undefined \def \showURL       {\relax}        \fi
\providecommand\bibfield[2]{#2}
\providecommand\bibinfo[2]{#2}
\providecommand\natexlab[1]{#1}
\providecommand\showeprint[2][]{arXiv:#2}

\bibitem[Agnese et~al\mbox{.}(2020)]%
        {agnese2020survey}
\bibfield{author}{\bibinfo{person}{Jorge Agnese}, \bibinfo{person}{Jonathan Herrera}, \bibinfo{person}{Haicheng Tao}, {and} \bibinfo{person}{Xingquan Zhu}.} \bibinfo{year}{2020}\natexlab{}.
\newblock \showarticletitle{A survey and taxonomy of adversarial neural networks for text-to-image synthesis}.
\newblock \bibinfo{journal}{\emph{Wiley Interdisciplinary Reviews: Data Mining and Knowledge Discovery}} \bibinfo{volume}{10}, \bibinfo{number}{4} (\bibinfo{year}{2020}), \bibinfo{pages}{e1345}.
\newblock


\bibitem[Ai et~al\mbox{.}(2023)]%
        {ai2023information}
\bibfield{author}{\bibinfo{person}{Qingyao Ai}, \bibinfo{person}{Ting Bai}, \bibinfo{person}{Zhao Cao}, \bibinfo{person}{Yi Chang}, \bibinfo{person}{Jiawei Chen}, \bibinfo{person}{Zhumin Chen}, \bibinfo{person}{Zhiyong Cheng}, \bibinfo{person}{Shoubin Dong}, \bibinfo{person}{Zhicheng Dou}, \bibinfo{person}{Fuli Feng}, {et~al\mbox{.}}} \bibinfo{year}{2023}\natexlab{}.
\newblock \showarticletitle{Information Retrieval Meets Large Language Models: A Strategic Report from Chinese IR Community}.
\newblock \bibinfo{journal}{\emph{AI Open}}  \bibinfo{volume}{4} (\bibinfo{year}{2023}), \bibinfo{pages}{80--90}.
\newblock


\bibitem[Alberts et~al\mbox{.}(2023)]%
        {alberts2023large}
\bibfield{author}{\bibinfo{person}{Ian~L Alberts}, \bibinfo{person}{Lorenzo Mercolli}, \bibinfo{person}{Thomas Pyka}, \bibinfo{person}{George Prenosil}, \bibinfo{person}{Kuangyu Shi}, \bibinfo{person}{Axel Rominger}, {and} \bibinfo{person}{Ali Afshar-Oromieh}.} \bibinfo{year}{2023}\natexlab{}.
\newblock \showarticletitle{Large language models (LLM) and ChatGPT: what will the impact on nuclear medicine be?}
\newblock \bibinfo{journal}{\emph{European journal of nuclear medicine and molecular imaging}} \bibinfo{volume}{50}, \bibinfo{number}{6} (\bibinfo{year}{2023}), \bibinfo{pages}{1549--1552}.
\newblock


\bibitem[Alemohammad et~al\mbox{.}(2023)]%
        {alemohammad2023self}
\bibfield{author}{\bibinfo{person}{Sina Alemohammad}, \bibinfo{person}{Josue Casco-Rodriguez}, \bibinfo{person}{Lorenzo Luzi}, \bibinfo{person}{Ahmed~Imtiaz Humayun}, \bibinfo{person}{Hossein Babaei}, \bibinfo{person}{Daniel LeJeune}, \bibinfo{person}{Ali Siahkoohi}, {and} \bibinfo{person}{Richard~G Baraniuk}.} \bibinfo{year}{2023}\natexlab{}.
\newblock \showarticletitle{Self-consuming generative models go mad}.
\newblock \bibinfo{journal}{\emph{arXiv preprint arXiv:2307.01850}} (\bibinfo{year}{2023}).
\newblock


\bibitem[Aslett et~al\mbox{.}(2023)]%
        {aslett2023online}
\bibfield{author}{\bibinfo{person}{Kevin Aslett}, \bibinfo{person}{Zeve Sanderson}, \bibinfo{person}{William Godel}, \bibinfo{person}{Nathaniel Persily}, \bibinfo{person}{Jonathan Nagler}, {and} \bibinfo{person}{Joshua~A Tucker}.} \bibinfo{year}{2023}\natexlab{}.
\newblock \showarticletitle{Online searches to evaluate misinformation can increase its perceived veracity}.
\newblock \bibinfo{journal}{\emph{Nature}} (\bibinfo{year}{2023}), \bibinfo{pages}{1--9}.
\newblock


\bibitem[Azzopardi et~al\mbox{.}(2003)]%
        {azzopardi2003investigating}
\bibfield{author}{\bibinfo{person}{Leif Azzopardi}, \bibinfo{person}{Mark Girolami}, {and} \bibinfo{person}{Keith Van~Risjbergen}.} \bibinfo{year}{2003}\natexlab{}.
\newblock \showarticletitle{Investigating the relationship between language model perplexity and IR precision-recall measures}. In \bibinfo{booktitle}{\emph{Proceedings of the 26th annual international ACM SIGIR conference on Research and development in informaion retrieval}}. \bibinfo{pages}{369--370}.
\newblock


\bibitem[Bang et~al\mbox{.}(2023)]%
        {bang2023multitask}
\bibfield{author}{\bibinfo{person}{Yejin Bang}, \bibinfo{person}{Samuel Cahyawijaya}, \bibinfo{person}{Nayeon Lee}, \bibinfo{person}{Wenliang Dai}, \bibinfo{person}{Dan Su}, \bibinfo{person}{Bryan Wilie}, \bibinfo{person}{Holy Lovenia}, \bibinfo{person}{Ziwei Ji}, \bibinfo{person}{Tiezheng Yu}, \bibinfo{person}{Willy Chung}, {et~al\mbox{.}}} \bibinfo{year}{2023}\natexlab{}.
\newblock \showarticletitle{A multitask, multilingual, multimodal evaluation of chatgpt on reasoning, hallucination, and interactivity}.
\newblock \bibinfo{journal}{\emph{arXiv preprint arXiv:2302.04023}} (\bibinfo{year}{2023}).
\newblock


\bibitem[Bengio et~al\mbox{.}(2023)]%
        {bengio2023managing}
\bibfield{author}{\bibinfo{person}{Yoshua Bengio}, \bibinfo{person}{Geoffrey Hinton}, \bibinfo{person}{Andrew Yao}, \bibinfo{person}{Dawn Song}, \bibinfo{person}{Pieter Abbeel}, \bibinfo{person}{Yuval~Noah Harari}, \bibinfo{person}{Ya-Qin Zhang}, \bibinfo{person}{Lan Xue}, \bibinfo{person}{Shai Shalev-Shwartz}, \bibinfo{person}{Gillian Hadfield}, {et~al\mbox{.}}} \bibinfo{year}{2023}\natexlab{}.
\newblock \showarticletitle{Managing ai risks in an era of rapid progress}.
\newblock \bibinfo{journal}{\emph{arXiv preprint arXiv:2310.17688}} (\bibinfo{year}{2023}).
\newblock


\bibitem[Betker et~al\mbox{.}(2023)]%
        {betker2023improving}
\bibfield{author}{\bibinfo{person}{James Betker}, \bibinfo{person}{Gabriel Goh}, \bibinfo{person}{Li Jing}, \bibinfo{person}{Tim Brooks}, \bibinfo{person}{Jianfeng Wang}, \bibinfo{person}{Linjie Li}, \bibinfo{person}{Long Ouyang}, \bibinfo{person}{Juntang Zhuang}, \bibinfo{person}{Joyce Lee}, \bibinfo{person}{Yufei Guo}, \bibinfo{person}{Wesam Manassra}, \bibinfo{person}{Prafulla Dhariwal}, \bibinfo{person}{Casey Chu}, {and} \bibinfo{person}{Yunxin Jiao}.} \bibinfo{year}{2023}\natexlab{}.
\newblock \showarticletitle{Improving Image Generation with Better Captions}.
\newblock  (\bibinfo{year}{2023}).
\newblock


\bibitem[Briesch et~al\mbox{.}(2023)]%
        {briesch2023large}
\bibfield{author}{\bibinfo{person}{Martin Briesch}, \bibinfo{person}{Dominik Sobania}, {and} \bibinfo{person}{Franz Rothlauf}.} \bibinfo{year}{2023}\natexlab{}.
\newblock \showarticletitle{Large Language Models Suffer From Their Own Output: An Analysis of the Self-Consuming Training Loop}.
\newblock \bibinfo{journal}{\emph{arXiv preprint arXiv:2311.16822}} (\bibinfo{year}{2023}).
\newblock


\bibitem[Bubeck et~al\mbox{.}(2023)]%
        {bubeck2023sparks}
\bibfield{author}{\bibinfo{person}{S{\'e}bastien Bubeck}, \bibinfo{person}{Varun Chandrasekaran}, \bibinfo{person}{Ronen Eldan}, \bibinfo{person}{Johannes Gehrke}, \bibinfo{person}{Eric Horvitz}, \bibinfo{person}{Ece Kamar}, \bibinfo{person}{Peter Lee}, \bibinfo{person}{Yin~Tat Lee}, \bibinfo{person}{Yuanzhi Li}, \bibinfo{person}{Scott Lundberg}, {et~al\mbox{.}}} \bibinfo{year}{2023}\natexlab{}.
\newblock \showarticletitle{Sparks of artificial general intelligence: Early experiments with gpt-4}.
\newblock \bibinfo{journal}{\emph{arXiv preprint arXiv:2303.12712}} (\bibinfo{year}{2023}).
\newblock


\bibitem[Cao et~al\mbox{.}(2023)]%
        {cao2023comprehensive}
\bibfield{author}{\bibinfo{person}{Yihan Cao}, \bibinfo{person}{Siyu Li}, \bibinfo{person}{Yixin Liu}, \bibinfo{person}{Zhiling Yan}, \bibinfo{person}{Yutong Dai}, \bibinfo{person}{Philip~S Yu}, {and} \bibinfo{person}{Lichao Sun}.} \bibinfo{year}{2023}\natexlab{}.
\newblock \showarticletitle{A comprehensive survey of ai-generated content (aigc): A history of generative ai from gan to chatgpt}.
\newblock \bibinfo{journal}{\emph{arXiv preprint arXiv:2303.04226}} (\bibinfo{year}{2023}).
\newblock


\bibitem[Chen and Shu(2023)]%
        {chen2023can}
\bibfield{author}{\bibinfo{person}{Canyu Chen} {and} \bibinfo{person}{Kai Shu}.} \bibinfo{year}{2023}\natexlab{}.
\newblock \showarticletitle{Can LLM-Generated Misinformation Be Detected?}
\newblock \bibinfo{journal}{\emph{arXiv preprint arXiv:2309.13788}} (\bibinfo{year}{2023}).
\newblock


\bibitem[Cho et~al\mbox{.}(2023)]%
        {cho2023discrete}
\bibfield{author}{\bibinfo{person}{Sukmin Cho}, \bibinfo{person}{Soyeong Jeong}, \bibinfo{person}{Jeongyeon Seo}, {and} \bibinfo{person}{Jong~C Park}.} \bibinfo{year}{2023}\natexlab{}.
\newblock \showarticletitle{Discrete Prompt Optimization via Constrained Generation for Zero-shot Re-ranker}.
\newblock \bibinfo{journal}{\emph{arXiv preprint arXiv:2305.13729}} (\bibinfo{year}{2023}).
\newblock


\bibitem[Dai et~al\mbox{.}(2023)]%
        {dai2023uncovering}
\bibfield{author}{\bibinfo{person}{Sunhao Dai}, \bibinfo{person}{Ninglu Shao}, \bibinfo{person}{Haiyuan Zhao}, \bibinfo{person}{Weijie Yu}, \bibinfo{person}{Zihua Si}, \bibinfo{person}{Chen Xu}, \bibinfo{person}{Zhongxiang Sun}, \bibinfo{person}{Xiao Zhang}, {and} \bibinfo{person}{Jun Xu}.} \bibinfo{year}{2023}\natexlab{}.
\newblock \showarticletitle{Uncovering ChatGPT's Capabilities in Recommender Systems}. In \bibinfo{booktitle}{\emph{Proceedings of the 17th ACM Conference on Recommender Systems}}.
\newblock


\bibitem[Dai et~al\mbox{.}(2022)]%
        {dai2022promptagator}
\bibfield{author}{\bibinfo{person}{Zhuyun Dai}, \bibinfo{person}{Vincent~Y Zhao}, \bibinfo{person}{Ji Ma}, \bibinfo{person}{Yi Luan}, \bibinfo{person}{Jianmo Ni}, \bibinfo{person}{Jing Lu}, \bibinfo{person}{Anton Bakalov}, \bibinfo{person}{Kelvin Guu}, \bibinfo{person}{Keith~B Hall}, {and} \bibinfo{person}{Ming-Wei Chang}.} \bibinfo{year}{2022}\natexlab{}.
\newblock \showarticletitle{Promptagator: Few-shot dense retrieval from 8 examples}.
\newblock \bibinfo{journal}{\emph{arXiv preprint arXiv:2209.11755}} (\bibinfo{year}{2022}).
\newblock


\bibitem[Del{\'e}tang et~al\mbox{.}(2023)]%
        {deletang2023language}
\bibfield{author}{\bibinfo{person}{Gr{\'e}goire Del{\'e}tang}, \bibinfo{person}{Anian Ruoss}, \bibinfo{person}{Paul-Ambroise Duquenne}, \bibinfo{person}{Elliot Catt}, \bibinfo{person}{Tim Genewein}, \bibinfo{person}{Christopher Mattern}, \bibinfo{person}{Jordi Grau-Moya}, \bibinfo{person}{Li~Kevin Wenliang}, \bibinfo{person}{Matthew Aitchison}, \bibinfo{person}{Laurent Orseau}, {et~al\mbox{.}}} \bibinfo{year}{2023}\natexlab{}.
\newblock \showarticletitle{Language modeling is compression}.
\newblock \bibinfo{journal}{\emph{arXiv preprint arXiv:2309.10668}} (\bibinfo{year}{2023}).
\newblock


\bibitem[Deshpande et~al\mbox{.}(2023)]%
        {deshpande2023toxicity}
\bibfield{author}{\bibinfo{person}{Ameet Deshpande}, \bibinfo{person}{Vishvak Murahari}, \bibinfo{person}{Tanmay Rajpurohit}, \bibinfo{person}{Ashwin Kalyan}, {and} \bibinfo{person}{Karthik Narasimhan}.} \bibinfo{year}{2023}\natexlab{}.
\newblock \showarticletitle{Toxicity in chatgpt: Analyzing persona-assigned language models}.
\newblock \bibinfo{journal}{\emph{arXiv preprint arXiv:2304.05335}} (\bibinfo{year}{2023}).
\newblock


\bibitem[Devlin et~al\mbox{.}(2019)]%
        {devlin-etal-2019-bert}
\bibfield{author}{\bibinfo{person}{Jacob Devlin}, \bibinfo{person}{Ming-Wei Chang}, \bibinfo{person}{Kenton Lee}, {and} \bibinfo{person}{Kristina Toutanova}.} \bibinfo{year}{2019}\natexlab{}.
\newblock \showarticletitle{{BERT}: Pre-training of Deep Bidirectional Transformers for Language Understanding}. In \bibinfo{booktitle}{\emph{Proceedings of the 2019 Conference of the North {A}merican Chapter of the Association for Computational Linguistics: Human Language Technologies}}. \bibinfo{pages}{4171--4186}.
\newblock


\bibitem[Fan et~al\mbox{.}(2023)]%
        {fan2023recommender}
\bibfield{author}{\bibinfo{person}{Wenqi Fan}, \bibinfo{person}{Zihuai Zhao}, \bibinfo{person}{Jiatong Li}, \bibinfo{person}{Yunqing Liu}, \bibinfo{person}{Xiaowei Mei}, \bibinfo{person}{Yiqi Wang}, \bibinfo{person}{Jiliang Tang}, {and} \bibinfo{person}{Qing Li}.} \bibinfo{year}{2023}\natexlab{}.
\newblock \showarticletitle{Recommender systems in the era of large language models (llms)}.
\newblock \bibinfo{journal}{\emph{arXiv preprint arXiv:2307.02046}} (\bibinfo{year}{2023}).
\newblock


\bibitem[Guo et~al\mbox{.}(2023)]%
        {guo2023close}
\bibfield{author}{\bibinfo{person}{Biyang Guo}, \bibinfo{person}{Xin Zhang}, \bibinfo{person}{Ziyuan Wang}, \bibinfo{person}{Minqi Jiang}, \bibinfo{person}{Jinran Nie}, \bibinfo{person}{Yuxuan Ding}, \bibinfo{person}{Jianwei Yue}, {and} \bibinfo{person}{Yupeng Wu}.} \bibinfo{year}{2023}\natexlab{}.
\newblock \showarticletitle{How Close is ChatGPT to Human Experts? Comparison Corpus, Evaluation, and Detection}.
\newblock \bibinfo{journal}{\emph{arXiv preprint arXiv:2301.07597}} (\bibinfo{year}{2023}).
\newblock


\bibitem[Guo et~al\mbox{.}(2022)]%
        {guo2022semantic}
\bibfield{author}{\bibinfo{person}{Jiafeng Guo}, \bibinfo{person}{Yinqiong Cai}, \bibinfo{person}{Yixing Fan}, \bibinfo{person}{Fei Sun}, \bibinfo{person}{Ruqing Zhang}, {and} \bibinfo{person}{Xueqi Cheng}.} \bibinfo{year}{2022}\natexlab{}.
\newblock \showarticletitle{Semantic models for the first-stage retrieval: A comprehensive review}.
\newblock \bibinfo{journal}{\emph{ACM Transactions on Information Systems (TOIS)}} \bibinfo{volume}{40}, \bibinfo{number}{4} (\bibinfo{year}{2022}), \bibinfo{pages}{1--42}.
\newblock


\bibitem[Guo et~al\mbox{.}(2020)]%
        {guo2020deep}
\bibfield{author}{\bibinfo{person}{Jiafeng Guo}, \bibinfo{person}{Yixing Fan}, \bibinfo{person}{Liang Pang}, \bibinfo{person}{Liu Yang}, \bibinfo{person}{Qingyao Ai}, \bibinfo{person}{Hamed Zamani}, \bibinfo{person}{Chen Wu}, \bibinfo{person}{W~Bruce Croft}, {and} \bibinfo{person}{Xueqi Cheng}.} \bibinfo{year}{2020}\natexlab{}.
\newblock \showarticletitle{A deep look into neural ranking models for information retrieval}.
\newblock \bibinfo{journal}{\emph{Information Processing \& Management}} \bibinfo{volume}{57}, \bibinfo{number}{6} (\bibinfo{year}{2020}), \bibinfo{pages}{102067}.
\newblock


\bibitem[Gy{\"o}ngyi et~al\mbox{.}(2004)]%
        {gyongyi2004combating}
\bibfield{author}{\bibinfo{person}{Zolt{\'a}n Gy{\"o}ngyi}, \bibinfo{person}{Hector Garcia-Molina}, {and} \bibinfo{person}{Jan Pedersen}.} \bibinfo{year}{2004}\natexlab{}.
\newblock \showarticletitle{Combating web spam with trustrank}. In \bibinfo{booktitle}{\emph{Proceedings of the Thirtieth international conference on Very large data bases-Volume 30}}. \bibinfo{pages}{576--587}.
\newblock


\bibitem[Hanley and Durumeric(2023)]%
        {hanley2023machine}
\bibfield{author}{\bibinfo{person}{Hans~WA Hanley} {and} \bibinfo{person}{Zakir Durumeric}.} \bibinfo{year}{2023}\natexlab{}.
\newblock \showarticletitle{Machine-Made Media: Monitoring the Mobilization of Machine-Generated Articles on Misinformation and Mainstream News Websites}.
\newblock \bibinfo{journal}{\emph{arXiv preprint arXiv:2305.09820}} (\bibinfo{year}{2023}).
\newblock


\bibitem[Hofst{\"a}tter et~al\mbox{.}(2021)]%
        {hofstatter2021efficiently}
\bibfield{author}{\bibinfo{person}{Sebastian Hofst{\"a}tter}, \bibinfo{person}{Sheng-Chieh Lin}, \bibinfo{person}{Jheng-Hong Yang}, \bibinfo{person}{Jimmy Lin}, {and} \bibinfo{person}{Allan Hanbury}.} \bibinfo{year}{2021}\natexlab{}.
\newblock \showarticletitle{Efficiently teaching an effective dense retriever with balanced topic aware sampling}. In \bibinfo{booktitle}{\emph{Proceedings of the 44th International ACM SIGIR Conference on Research and Development in Information Retrieval}}. \bibinfo{pages}{113--122}.
\newblock


\bibitem[Huang et~al\mbox{.}(2023)]%
        {huang2023finbert}
\bibfield{author}{\bibinfo{person}{Allen~H Huang}, \bibinfo{person}{Hui Wang}, {and} \bibinfo{person}{Yi Yang}.} \bibinfo{year}{2023}\natexlab{}.
\newblock \showarticletitle{FinBERT: A large language model for extracting information from financial text}.
\newblock \bibinfo{journal}{\emph{Contemporary Accounting Research}} \bibinfo{volume}{40}, \bibinfo{number}{2} (\bibinfo{year}{2023}), \bibinfo{pages}{806--841}.
\newblock


\bibitem[Izacard et~al\mbox{.}(2021)]%
        {izacard2021unsupervised}
\bibfield{author}{\bibinfo{person}{Gautier Izacard}, \bibinfo{person}{Mathilde Caron}, \bibinfo{person}{Lucas Hosseini}, \bibinfo{person}{Sebastian Riedel}, \bibinfo{person}{Piotr Bojanowski}, \bibinfo{person}{Armand Joulin}, {and} \bibinfo{person}{Edouard Grave}.} \bibinfo{year}{2021}\natexlab{}.
\newblock \showarticletitle{Unsupervised dense information retrieval with contrastive learning}.
\newblock \bibinfo{journal}{\emph{arXiv preprint arXiv:2112.09118}} (\bibinfo{year}{2021}).
\newblock


\bibitem[Izacard et~al\mbox{.}(2022)]%
        {izacard2022few}
\bibfield{author}{\bibinfo{person}{Gautier Izacard}, \bibinfo{person}{Patrick Lewis}, \bibinfo{person}{Maria Lomeli}, \bibinfo{person}{Lucas Hosseini}, \bibinfo{person}{Fabio Petroni}, \bibinfo{person}{Timo Schick}, \bibinfo{person}{Jane Dwivedi-Yu}, \bibinfo{person}{Armand Joulin}, \bibinfo{person}{Sebastian Riedel}, {and} \bibinfo{person}{Edouard Grave}.} \bibinfo{year}{2022}\natexlab{}.
\newblock \showarticletitle{Few-shot learning with retrieval augmented language models}.
\newblock \bibinfo{journal}{\emph{arXiv preprint arXiv:2208.03299}} (\bibinfo{year}{2022}).
\newblock


\bibitem[Jiang et~al\mbox{.}(2023)]%
        {jiang2023disinformation}
\bibfield{author}{\bibinfo{person}{Bohan Jiang}, \bibinfo{person}{Zhen Tan}, \bibinfo{person}{Ayushi Nirmal}, {and} \bibinfo{person}{Huan Liu}.} \bibinfo{year}{2023}\natexlab{}.
\newblock \showarticletitle{Disinformation Detection: An Evolving Challenge in the Age of LLMs}.
\newblock \bibinfo{journal}{\emph{arXiv preprint arXiv:2309.15847}} (\bibinfo{year}{2023}).
\newblock


\bibitem[Klema and Laub(1980)]%
        {klema1980singular}
\bibfield{author}{\bibinfo{person}{Virginia Klema} {and} \bibinfo{person}{Alan Laub}.} \bibinfo{year}{1980}\natexlab{}.
\newblock \showarticletitle{The singular value decomposition: Its computation and some applications}.
\newblock \bibinfo{journal}{\emph{IEEE Transactions on automatic control}} \bibinfo{volume}{25}, \bibinfo{number}{2} (\bibinfo{year}{1980}), \bibinfo{pages}{164--176}.
\newblock


\bibitem[Kwiatkowski et~al\mbox{.}(2019)]%
        {kwiatkowski2019natural}
\bibfield{author}{\bibinfo{person}{Tom Kwiatkowski}, \bibinfo{person}{Jennimaria Palomaki}, \bibinfo{person}{Olivia Redfield}, \bibinfo{person}{Michael Collins}, \bibinfo{person}{Ankur Parikh}, \bibinfo{person}{Chris Alberti}, \bibinfo{person}{Danielle Epstein}, \bibinfo{person}{Illia Polosukhin}, \bibinfo{person}{Jacob Devlin}, \bibinfo{person}{Kenton Lee}, {et~al\mbox{.}}} \bibinfo{year}{2019}\natexlab{}.
\newblock \showarticletitle{Natural questions: a benchmark for question answering research}.
\newblock \bibinfo{journal}{\emph{Transactions of the Association for Computational Linguistics}}  \bibinfo{volume}{7} (\bibinfo{year}{2019}), \bibinfo{pages}{453--466}.
\newblock


\bibitem[Li(2022)]%
        {li2022learning}
\bibfield{author}{\bibinfo{person}{Hang Li}.} \bibinfo{year}{2022}\natexlab{}.
\newblock \bibinfo{booktitle}{\emph{Learning to rank for information retrieval and natural language processing}}.
\newblock \bibinfo{publisher}{Springer Nature}.
\newblock


\bibitem[Liu et~al\mbox{.}(2009)]%
        {liu2009learning}
\bibfield{author}{\bibinfo{person}{Tie-Yan Liu} {et~al\mbox{.}}} \bibinfo{year}{2009}\natexlab{}.
\newblock \showarticletitle{Learning to rank for information retrieval}.
\newblock \bibinfo{journal}{\emph{Foundations and Trends{\textregistered} in Information Retrieval}} \bibinfo{volume}{3}, \bibinfo{number}{3} (\bibinfo{year}{2009}), \bibinfo{pages}{225--331}.
\newblock


\bibitem[Liu et~al\mbox{.}(2019)]%
        {liu2019roberta}
\bibfield{author}{\bibinfo{person}{Yinhan Liu}, \bibinfo{person}{Myle Ott}, \bibinfo{person}{Naman Goyal}, \bibinfo{person}{Jingfei Du}, \bibinfo{person}{Mandar Joshi}, \bibinfo{person}{Danqi Chen}, \bibinfo{person}{Omer Levy}, \bibinfo{person}{Mike Lewis}, \bibinfo{person}{Luke Zettlemoyer}, {and} \bibinfo{person}{Veselin Stoyanov}.} \bibinfo{year}{2019}\natexlab{}.
\newblock \showarticletitle{Roberta: A robustly optimized bert pretraining approach}.
\newblock \bibinfo{journal}{\emph{arXiv preprint arXiv:1907.11692}} (\bibinfo{year}{2019}).
\newblock


\bibitem[Manning(2009)]%
        {manning2009introduction}
\bibfield{author}{\bibinfo{person}{Christopher~D Manning}.} \bibinfo{year}{2009}\natexlab{}.
\newblock \bibinfo{booktitle}{\emph{An introduction to information retrieval}}.
\newblock \bibinfo{publisher}{Cambridge university press}.
\newblock


\bibitem[Nogueira et~al\mbox{.}(2020)]%
        {nogueira2020document}
\bibfield{author}{\bibinfo{person}{Rodrigo Nogueira}, \bibinfo{person}{Zhiying Jiang}, {and} \bibinfo{person}{Jimmy Lin}.} \bibinfo{year}{2020}\natexlab{}.
\newblock \showarticletitle{Document ranking with a pretrained sequence-to-sequence model}.
\newblock \bibinfo{journal}{\emph{arXiv preprint arXiv:2003.06713}} (\bibinfo{year}{2020}).
\newblock


\bibitem[Nunes et~al\mbox{.}(2023)]%
        {Nunes2023EvaluatingGA}
\bibfield{author}{\bibinfo{person}{Desnes Nunes}, \bibinfo{person}{Ricardo Primi}, \bibinfo{person}{Ramon Pires}, \bibinfo{person}{Roberto de Alencar~Lotufo}, {and} \bibinfo{person}{Rodrigo Nogueira}.} \bibinfo{year}{2023}\natexlab{}.
\newblock \showarticletitle{Evaluating GPT-3.5 and GPT-4 Models on Brazilian University Admission Exams}.
\newblock \bibinfo{journal}{\emph{ArXiv}}  \bibinfo{volume}{abs/2303.17003} (\bibinfo{year}{2023}).
\newblock


\bibitem[Pan et~al\mbox{.}(2023)]%
        {pan2023risk}
\bibfield{author}{\bibinfo{person}{Yikang Pan}, \bibinfo{person}{Liangming Pan}, \bibinfo{person}{Wenhu Chen}, \bibinfo{person}{Preslav Nakov}, \bibinfo{person}{Min-Yen Kan}, {and} \bibinfo{person}{William~Yang Wang}.} \bibinfo{year}{2023}\natexlab{}.
\newblock \showarticletitle{On the Risk of Misinformation Pollution with Large Language Models}.
\newblock \bibinfo{journal}{\emph{arXiv preprint arXiv:2305.13661}} (\bibinfo{year}{2023}).
\newblock


\bibitem[Raffel et~al\mbox{.}(2020)]%
        {raffel2020exploring}
\bibfield{author}{\bibinfo{person}{Colin Raffel}, \bibinfo{person}{Noam Shazeer}, \bibinfo{person}{Adam Roberts}, \bibinfo{person}{Katherine Lee}, \bibinfo{person}{Sharan Narang}, \bibinfo{person}{Michael Matena}, \bibinfo{person}{Yanqi Zhou}, \bibinfo{person}{Wei Li}, {and} \bibinfo{person}{Peter~J Liu}.} \bibinfo{year}{2020}\natexlab{}.
\newblock \showarticletitle{Exploring the limits of transfer learning with a unified text-to-text transformer}.
\newblock \bibinfo{journal}{\emph{The Journal of Machine Learning Research}} \bibinfo{volume}{21}, \bibinfo{number}{1} (\bibinfo{year}{2020}), \bibinfo{pages}{5485--5551}.
\newblock


\bibitem[Robertson et~al\mbox{.}(2009)]%
        {robertson2009probabilistic}
\bibfield{author}{\bibinfo{person}{Stephen Robertson}, \bibinfo{person}{Hugo Zaragoza}, {et~al\mbox{.}}} \bibinfo{year}{2009}\natexlab{}.
\newblock \showarticletitle{The probabilistic relevance framework: BM25 and beyond}.
\newblock \bibinfo{journal}{\emph{Foundations and Trends{\textregistered} in Information Retrieval}} \bibinfo{volume}{3}, \bibinfo{number}{4} (\bibinfo{year}{2009}), \bibinfo{pages}{333--389}.
\newblock


\bibitem[Sadasivan et~al\mbox{.}(2023)]%
        {Sadasivan2023CanAT}
\bibfield{author}{\bibinfo{person}{Vinu~Sankar Sadasivan}, \bibinfo{person}{Aounon Kumar}, \bibinfo{person}{S. Balasubramanian}, \bibinfo{person}{Wenxiao Wang}, {and} \bibinfo{person}{Soheil Feizi}.} \bibinfo{year}{2023}\natexlab{}.
\newblock \showarticletitle{Can AI-Generated Text be Reliably Detected?}
\newblock \bibinfo{journal}{\emph{ArXiv}}  \bibinfo{volume}{abs/2303.11156} (\bibinfo{year}{2023}).
\newblock
\urldef\tempurl%
\url{https://api.semanticscholar.org/CorpusID:257631570}
\showURL{%
\tempurl}


\bibitem[Shi et~al\mbox{.}(2023)]%
        {shi2023replug}
\bibfield{author}{\bibinfo{person}{Weijia Shi}, \bibinfo{person}{Sewon Min}, \bibinfo{person}{Michihiro Yasunaga}, \bibinfo{person}{Minjoon Seo}, \bibinfo{person}{Rich James}, \bibinfo{person}{Mike Lewis}, \bibinfo{person}{Luke Zettlemoyer}, {and} \bibinfo{person}{Wen-tau Yih}.} \bibinfo{year}{2023}\natexlab{}.
\newblock \showarticletitle{Replug: Retrieval-augmented black-box language models}.
\newblock \bibinfo{journal}{\emph{arXiv preprint arXiv:2301.12652}} (\bibinfo{year}{2023}).
\newblock


\bibitem[Shumailov et~al\mbox{.}(2023)]%
        {shumailov2023model}
\bibfield{author}{\bibinfo{person}{Ilia Shumailov}, \bibinfo{person}{Zakhar Shumaylov}, \bibinfo{person}{Yiren Zhao}, \bibinfo{person}{Yarin Gal}, \bibinfo{person}{Nicolas Papernot}, {and} \bibinfo{person}{Ross Anderson}.} \bibinfo{year}{2023}\natexlab{}.
\newblock \showarticletitle{Model dementia: Generated data makes models forget}.
\newblock \bibinfo{journal}{\emph{arXiv e-prints}} (\bibinfo{year}{2023}), \bibinfo{pages}{arXiv--2305}.
\newblock


\bibitem[Singhal et~al\mbox{.}(2001)]%
        {singhal2001modern}
\bibfield{author}{\bibinfo{person}{Amit Singhal} {et~al\mbox{.}}} \bibinfo{year}{2001}\natexlab{}.
\newblock \showarticletitle{Modern information retrieval: A brief overview}.
\newblock \bibinfo{journal}{\emph{IEEE Data Eng. Bull.}} \bibinfo{volume}{24}, \bibinfo{number}{4} (\bibinfo{year}{2001}), \bibinfo{pages}{35--43}.
\newblock


\bibitem[Sparck~Jones(1972)]%
        {sparck1972statistical}
\bibfield{author}{\bibinfo{person}{Karen Sparck~Jones}.} \bibinfo{year}{1972}\natexlab{}.
\newblock \showarticletitle{A statistical interpretation of term specificity and its application in retrieval}.
\newblock \bibinfo{journal}{\emph{Journal of documentation}} \bibinfo{volume}{28}, \bibinfo{number}{1} (\bibinfo{year}{1972}), \bibinfo{pages}{11--21}.
\newblock


\bibitem[Spitale et~al\mbox{.}(2023)]%
        {spitale2023ai}
\bibfield{author}{\bibinfo{person}{Giovanni Spitale}, \bibinfo{person}{Nikola Biller-Andorno}, {and} \bibinfo{person}{Federico Germani}.} \bibinfo{year}{2023}\natexlab{}.
\newblock \showarticletitle{AI model GPT-3 (dis) informs us better than humans}.
\newblock \bibinfo{journal}{\emph{arXiv preprint arXiv:2301.11924}} (\bibinfo{year}{2023}).
\newblock


\bibitem[Srinivasan et~al\mbox{.}(2022)]%
        {srinivasan2022quill}
\bibfield{author}{\bibinfo{person}{Krishna Srinivasan}, \bibinfo{person}{Karthik Raman}, \bibinfo{person}{Anupam Samanta}, \bibinfo{person}{Lingrui Liao}, \bibinfo{person}{Luca Bertelli}, {and} \bibinfo{person}{Michael Bendersky}.} \bibinfo{year}{2022}\natexlab{}.
\newblock \showarticletitle{QUILL: Query Intent with Large Language Models using Retrieval Augmentation and Multi-stage Distillation}. In \bibinfo{booktitle}{\emph{Proceedings of the 2022 Conference on Empirical Methods in Natural Language Processing: Industry Track}}. \bibinfo{pages}{492--501}.
\newblock


\bibitem[Su et~al\mbox{.}(2023)]%
        {Su2023FakeND}
\bibfield{author}{\bibinfo{person}{Jinyan Su}, \bibinfo{person}{Terry~Yue Zhuo}, \bibinfo{person}{Jonibek Mansurov}, \bibinfo{person}{Di Wang}, {and} \bibinfo{person}{Preslav Nakov}.} \bibinfo{year}{2023}\natexlab{}.
\newblock \showarticletitle{Fake News Detectors are Biased against Texts Generated by Large Language Models}.
\newblock \bibinfo{journal}{\emph{arXiv preprint arxiv:2309.08674}} (\bibinfo{year}{2023}).
\newblock


\bibitem[Sun et~al\mbox{.}(2023)]%
        {sun2023chatgpt}
\bibfield{author}{\bibinfo{person}{Weiwei Sun}, \bibinfo{person}{Lingyong Yan}, \bibinfo{person}{Xinyu Ma}, \bibinfo{person}{Pengjie Ren}, \bibinfo{person}{Dawei Yin}, {and} \bibinfo{person}{Zhaochun Ren}.} \bibinfo{year}{2023}\natexlab{}.
\newblock \showarticletitle{Is ChatGPT Good at Search? Investigating Large Language Models as Re-Ranking Agent}.
\newblock \bibinfo{journal}{\emph{arXiv preprint arXiv:2304.09542}} (\bibinfo{year}{2023}).
\newblock


\bibitem[Thakur et~al\mbox{.}(2023)]%
        {thakur-etal-2023-language}
\bibfield{author}{\bibinfo{person}{Himanshu Thakur}, \bibinfo{person}{Atishay Jain}, \bibinfo{person}{Praneetha Vaddamanu}, \bibinfo{person}{Paul~Pu Liang}, {and} \bibinfo{person}{Louis-Philippe Morency}.} \bibinfo{year}{2023}\natexlab{}.
\newblock \showarticletitle{Language Models Get a Gender Makeover: Mitigating Gender Bias with Few-Shot Data Interventions}. In \bibinfo{booktitle}{\emph{Proceedings of the 61st Annual Meeting of the Association for Computational Linguistics}}. \bibinfo{pages}{340--351}.
\newblock


\bibitem[Thakur et~al\mbox{.}(2021)]%
        {thakur2021beir}
\bibfield{author}{\bibinfo{person}{Nandan Thakur}, \bibinfo{person}{Nils Reimers}, \bibinfo{person}{Andreas R{\"u}ckl{\'e}}, \bibinfo{person}{Abhishek Srivastava}, {and} \bibinfo{person}{Iryna Gurevych}.} \bibinfo{year}{2021}\natexlab{}.
\newblock \showarticletitle{{BEIR}: A Heterogeneous Benchmark for Zero-shot Evaluation of Information Retrieval Models}. In \bibinfo{booktitle}{\emph{Thirty-fifth Conference on Neural Information Processing Systems Datasets and Benchmarks Track}}.
\newblock


\bibitem[Thirunavukarasu et~al\mbox{.}(2023)]%
        {thirunavukarasu2023large}
\bibfield{author}{\bibinfo{person}{Arun~James Thirunavukarasu}, \bibinfo{person}{Darren Shu~Jeng Ting}, \bibinfo{person}{Kabilan Elangovan}, \bibinfo{person}{Laura Gutierrez}, \bibinfo{person}{Ting~Fang Tan}, {and} \bibinfo{person}{Daniel Shu~Wei Ting}.} \bibinfo{year}{2023}\natexlab{}.
\newblock \showarticletitle{Large language models in medicine}.
\newblock \bibinfo{journal}{\emph{Nature medicine}} \bibinfo{volume}{29}, \bibinfo{number}{8} (\bibinfo{year}{2023}), \bibinfo{pages}{1930--1940}.
\newblock


\bibitem[Touvron et~al\mbox{.}(2023)]%
        {touvron2023llama}
\bibfield{author}{\bibinfo{person}{Hugo Touvron}, \bibinfo{person}{Louis Martin}, \bibinfo{person}{Kevin Stone}, \bibinfo{person}{Peter Albert}, \bibinfo{person}{Amjad Almahairi}, \bibinfo{person}{Yasmine Babaei}, \bibinfo{person}{Nikolay Bashlykov}, \bibinfo{person}{Soumya Batra}, \bibinfo{person}{Prajjwal Bhargava}, \bibinfo{person}{Shruti Bhosale}, {et~al\mbox{.}}} \bibinfo{year}{2023}\natexlab{}.
\newblock \showarticletitle{Llama 2: Open foundation and fine-tuned chat models}.
\newblock \bibinfo{journal}{\emph{arXiv preprint arXiv:2307.09288}} (\bibinfo{year}{2023}).
\newblock


\bibitem[Van~der Maaten and Hinton(2008)]%
        {van2008visualizing}
\bibfield{author}{\bibinfo{person}{Laurens Van~der Maaten} {and} \bibinfo{person}{Geoffrey Hinton}.} \bibinfo{year}{2008}\natexlab{}.
\newblock \showarticletitle{Visualizing data using t-SNE.}
\newblock \bibinfo{journal}{\emph{Journal of machine learning research}} \bibinfo{volume}{9}, \bibinfo{number}{11} (\bibinfo{year}{2008}).
\newblock


\bibitem[Vaswani et~al\mbox{.}(2017)]%
        {vaswani2017attention}
\bibfield{author}{\bibinfo{person}{Ashish Vaswani}, \bibinfo{person}{Noam Shazeer}, \bibinfo{person}{Niki Parmar}, \bibinfo{person}{Jakob Uszkoreit}, \bibinfo{person}{Llion Jones}, \bibinfo{person}{Aidan~N Gomez}, \bibinfo{person}{{\L}ukasz Kaiser}, {and} \bibinfo{person}{Illia Polosukhin}.} \bibinfo{year}{2017}\natexlab{}.
\newblock \showarticletitle{Attention is all you need}.
\newblock \bibinfo{journal}{\emph{Advances in neural information processing systems}}  \bibinfo{volume}{30} (\bibinfo{year}{2017}).
\newblock


\bibitem[Wadden et~al\mbox{.}(2020)]%
        {wadden2020fact}
\bibfield{author}{\bibinfo{person}{David Wadden}, \bibinfo{person}{Shanchuan Lin}, \bibinfo{person}{Kyle Lo}, \bibinfo{person}{Lucy~Lu Wang}, \bibinfo{person}{Madeleine van Zuylen}, \bibinfo{person}{Arman Cohan}, {and} \bibinfo{person}{Hannaneh Hajishirzi}.} \bibinfo{year}{2020}\natexlab{}.
\newblock \showarticletitle{Fact or fiction: Verifying scientific claims}.
\newblock \bibinfo{journal}{\emph{arXiv preprint arXiv:2004.14974}} (\bibinfo{year}{2020}).
\newblock


\bibitem[Wang and Cho(2019)]%
        {wang2019bert}
\bibfield{author}{\bibinfo{person}{Alex Wang} {and} \bibinfo{person}{Kyunghyun Cho}.} \bibinfo{year}{2019}\natexlab{}.
\newblock \showarticletitle{BERT has a mouth, and it must speak: BERT as a Markov random field language model}.
\newblock \bibinfo{journal}{\emph{arXiv preprint arXiv:1902.04094}} (\bibinfo{year}{2019}).
\newblock


\bibitem[Wang et~al\mbox{.}(2019)]%
        {wang2019language}
\bibfield{author}{\bibinfo{person}{Chenguang Wang}, \bibinfo{person}{Mu Li}, {and} \bibinfo{person}{Alexander~J Smola}.} \bibinfo{year}{2019}\natexlab{}.
\newblock \showarticletitle{Language models with transformers}.
\newblock \bibinfo{journal}{\emph{arXiv preprint arXiv:1904.09408}} (\bibinfo{year}{2019}).
\newblock


\bibitem[Wang et~al\mbox{.}(2023a)]%
        {wang2023query2doc}
\bibfield{author}{\bibinfo{person}{Liang Wang}, \bibinfo{person}{Nan Yang}, {and} \bibinfo{person}{Furu Wei}.} \bibinfo{year}{2023}\natexlab{a}.
\newblock \showarticletitle{Query2doc: Query Expansion with Large Language Models}.
\newblock \bibinfo{journal}{\emph{arXiv preprint arXiv:2303.07678}} (\bibinfo{year}{2023}).
\newblock


\bibitem[Wang et~al\mbox{.}(2023b)]%
        {wang2023security}
\bibfield{author}{\bibinfo{person}{Tao Wang}, \bibinfo{person}{Yushu Zhang}, \bibinfo{person}{Shuren Qi}, \bibinfo{person}{Ruoyu Zhao}, \bibinfo{person}{Zhihua Xia}, {and} \bibinfo{person}{Jian Weng}.} \bibinfo{year}{2023}\natexlab{b}.
\newblock \showarticletitle{Security and privacy on generative data in aigc: A survey}.
\newblock \bibinfo{journal}{\emph{arXiv preprint arXiv:2309.09435}} (\bibinfo{year}{2023}).
\newblock


\bibitem[Wang et~al\mbox{.}(2020)]%
        {wang2020minilm}
\bibfield{author}{\bibinfo{person}{Wenhui Wang}, \bibinfo{person}{Furu Wei}, \bibinfo{person}{Li Dong}, \bibinfo{person}{Hangbo Bao}, \bibinfo{person}{Nan Yang}, {and} \bibinfo{person}{Ming Zhou}.} \bibinfo{year}{2020}\natexlab{}.
\newblock \showarticletitle{Minilm: Deep self-attention distillation for task-agnostic compression of pre-trained transformers}.
\newblock \bibinfo{journal}{\emph{Advances in Neural Information Processing Systems}}  \bibinfo{volume}{33} (\bibinfo{year}{2020}), \bibinfo{pages}{5776--5788}.
\newblock


\bibitem[Wang et~al\mbox{.}(2021)]%
        {wang2021reinforcing}
\bibfield{author}{\bibinfo{person}{Xiting Wang}, \bibinfo{person}{Xinwei Gu}, \bibinfo{person}{Jie Cao}, \bibinfo{person}{Zihua Zhao}, \bibinfo{person}{Yulan Yan}, \bibinfo{person}{Bhuvan Middha}, {and} \bibinfo{person}{Xing Xie}.} \bibinfo{year}{2021}\natexlab{}.
\newblock \showarticletitle{Reinforcing pretrained models for generating attractive text advertisements}. In \bibinfo{booktitle}{\emph{Proceedings of the 27th ACM SIGKDD Conference on Knowledge Discovery \& Data Mining}}. \bibinfo{pages}{3697--3707}.
\newblock


\bibitem[Wei et~al\mbox{.}(2022)]%
        {wei2022emergent}
\bibfield{author}{\bibinfo{person}{Jason Wei}, \bibinfo{person}{Yi Tay}, \bibinfo{person}{Rishi Bommasani}, \bibinfo{person}{Colin Raffel}, \bibinfo{person}{Barret Zoph}, \bibinfo{person}{Sebastian Borgeaud}, \bibinfo{person}{Dani Yogatama}, \bibinfo{person}{Maarten Bosma}, \bibinfo{person}{Denny Zhou}, \bibinfo{person}{Donald Metzler}, {et~al\mbox{.}}} \bibinfo{year}{2022}\natexlab{}.
\newblock \showarticletitle{Emergent abilities of large language models}.
\newblock \bibinfo{journal}{\emph{arXiv preprint arXiv:2206.07682}} (\bibinfo{year}{2022}).
\newblock


\bibitem[Wu et~al\mbox{.}(2023a)]%
        {wu2023ai}
\bibfield{author}{\bibinfo{person}{Jiayang Wu}, \bibinfo{person}{Wensheng Gan}, \bibinfo{person}{Zefeng Chen}, \bibinfo{person}{Shicheng Wan}, {and} \bibinfo{person}{Hong Lin}.} \bibinfo{year}{2023}\natexlab{a}.
\newblock \showarticletitle{Ai-generated content (aigc): A survey}.
\newblock \bibinfo{journal}{\emph{arXiv preprint arXiv:2304.06632}} (\bibinfo{year}{2023}).
\newblock


\bibitem[Wu et~al\mbox{.}(2023b)]%
        {wu2023bloomberggpt}
\bibfield{author}{\bibinfo{person}{Shijie Wu}, \bibinfo{person}{Ozan Irsoy}, \bibinfo{person}{Steven Lu}, \bibinfo{person}{Vadim Dabravolski}, \bibinfo{person}{Mark Dredze}, \bibinfo{person}{Sebastian Gehrmann}, \bibinfo{person}{Prabhanjan Kambadur}, \bibinfo{person}{David Rosenberg}, {and} \bibinfo{person}{Gideon Mann}.} \bibinfo{year}{2023}\natexlab{b}.
\newblock \showarticletitle{Bloomberggpt: A large language model for finance}.
\newblock \bibinfo{journal}{\emph{arXiv preprint arXiv:2303.17564}} (\bibinfo{year}{2023}).
\newblock


\bibitem[Xiong et~al\mbox{.}(2020)]%
        {xiong2020approximate}
\bibfield{author}{\bibinfo{person}{Lee Xiong}, \bibinfo{person}{Chenyan Xiong}, \bibinfo{person}{Ye Li}, \bibinfo{person}{Kwok-Fung Tang}, \bibinfo{person}{Jialin Liu}, \bibinfo{person}{Paul Bennett}, \bibinfo{person}{Junaid Ahmed}, {and} \bibinfo{person}{Arnold Overwijk}.} \bibinfo{year}{2020}\natexlab{}.
\newblock \showarticletitle{Approximate nearest neighbor negative contrastive learning for dense text retrieval}.
\newblock \bibinfo{journal}{\emph{arXiv preprint arXiv:2007.00808}} (\bibinfo{year}{2020}).
\newblock


\bibitem[Xu and Li(2007)]%
        {xu2007adarank}
\bibfield{author}{\bibinfo{person}{Jun Xu} {and} \bibinfo{person}{Hang Li}.} \bibinfo{year}{2007}\natexlab{}.
\newblock \showarticletitle{Adarank: a boosting algorithm for information retrieval}. In \bibinfo{booktitle}{\emph{Proceedings of the 30th annual international ACM SIGIR conference on Research and development in information retrieval}}. \bibinfo{pages}{391--398}.
\newblock


\bibitem[Xu et~al\mbox{.}(2022)]%
        {xu2022match}
\bibfield{author}{\bibinfo{person}{Shicheng Xu}, \bibinfo{person}{Liang Pang}, \bibinfo{person}{Huawei Shen}, {and} \bibinfo{person}{Xueqi Cheng}.} \bibinfo{year}{2022}\natexlab{}.
\newblock \showarticletitle{Match-Prompt: Improving Multi-task Generalization Ability for Neural Text Matching via Prompt Learning}. In \bibinfo{booktitle}{\emph{Proceedings of the 31st ACM International Conference on Information \& Knowledge Management}}. \bibinfo{pages}{2290--2300}.
\newblock


\bibitem[Xu et~al\mbox{.}(2023)]%
        {xu-etal-2023-berm}
\bibfield{author}{\bibinfo{person}{Shicheng Xu}, \bibinfo{person}{Liang Pang}, \bibinfo{person}{Huawei Shen}, {and} \bibinfo{person}{Xueqi Cheng}.} \bibinfo{year}{2023}\natexlab{}.
\newblock \showarticletitle{{BERM}: Training the Balanced and Extractable Representation for Matching to Improve Generalization Ability of Dense Retrieval}. In \bibinfo{booktitle}{\emph{Proceedings of the 61st Annual Meeting of the Association for Computational Linguistics}}. \bibinfo{pages}{6620--6635}.
\newblock


\bibitem[Yang et~al\mbox{.}(2023)]%
        {yang2023harnessing}
\bibfield{author}{\bibinfo{person}{Jingfeng Yang}, \bibinfo{person}{Hongye Jin}, \bibinfo{person}{Ruixiang Tang}, \bibinfo{person}{Xiaotian Han}, \bibinfo{person}{Qizhang Feng}, \bibinfo{person}{Haoming Jiang}, \bibinfo{person}{Bing Yin}, {and} \bibinfo{person}{Xia Hu}.} \bibinfo{year}{2023}\natexlab{}.
\newblock \showarticletitle{Harnessing the power of llms in practice: A survey on chatgpt and beyond}.
\newblock \bibinfo{journal}{\emph{arXiv preprint arXiv:2304.13712}} (\bibinfo{year}{2023}).
\newblock


\bibitem[Yates et~al\mbox{.}(2021)]%
        {yates2021pretrained}
\bibfield{author}{\bibinfo{person}{Andrew Yates}, \bibinfo{person}{Rodrigo Nogueira}, {and} \bibinfo{person}{Jimmy Lin}.} \bibinfo{year}{2021}\natexlab{}.
\newblock \showarticletitle{Pretrained transformers for text ranking: BERT and beyond}. In \bibinfo{booktitle}{\emph{Proceedings of the 14th ACM International Conference on web search and data mining}}. \bibinfo{pages}{1154--1156}.
\newblock


\bibitem[Zhao et~al\mbox{.}(2023a)]%
        {zhao2022dense}
\bibfield{author}{\bibinfo{person}{Wayne~Xin Zhao}, \bibinfo{person}{Jing Liu}, \bibinfo{person}{Ruiyang Ren}, {and} \bibinfo{person}{Ji-Rong Wen}.} \bibinfo{year}{2023}\natexlab{a}.
\newblock \showarticletitle{Dense Text Retrieval based on Pretrained Language Models: A Survey}.
\newblock \bibinfo{journal}{\emph{ACM Trans. Inf. Syst.}} (\bibinfo{date}{dec} \bibinfo{year}{2023}).
\newblock


\bibitem[Zhao et~al\mbox{.}(2023b)]%
        {zhao2023survey}
\bibfield{author}{\bibinfo{person}{Wayne~Xin Zhao}, \bibinfo{person}{Kun Zhou}, \bibinfo{person}{Junyi Li}, \bibinfo{person}{Tianyi Tang}, \bibinfo{person}{Xiaolei Wang}, \bibinfo{person}{Yupeng Hou}, \bibinfo{person}{Yingqian Min}, \bibinfo{person}{Beichen Zhang}, \bibinfo{person}{Junjie Zhang}, \bibinfo{person}{Zican Dong}, {et~al\mbox{.}}} \bibinfo{year}{2023}\natexlab{b}.
\newblock \showarticletitle{A survey of large language models}.
\newblock \bibinfo{journal}{\emph{arXiv preprint arXiv:2303.18223}} (\bibinfo{year}{2023}).
\newblock


\bibitem[Zhu et~al\mbox{.}(2023)]%
        {zhu2023large}
\bibfield{author}{\bibinfo{person}{Yutao Zhu}, \bibinfo{person}{Huaying Yuan}, \bibinfo{person}{Shuting Wang}, \bibinfo{person}{Jiongnan Liu}, \bibinfo{person}{Wenhan Liu}, \bibinfo{person}{Chenlong Deng}, \bibinfo{person}{Zhicheng Dou}, {and} \bibinfo{person}{Ji-Rong Wen}.} \bibinfo{year}{2023}\natexlab{}.
\newblock \showarticletitle{Large language models for information retrieval: A survey}.
\newblock \bibinfo{journal}{\emph{arXiv preprint arXiv:2308.07107}} (\bibinfo{year}{2023}).
\newblock


\bibitem[Zhuo et~al\mbox{.}(2023)]%
        {zhuo2023exploring}
\bibfield{author}{\bibinfo{person}{Terry~Yue Zhuo}, \bibinfo{person}{Yujin Huang}, \bibinfo{person}{Chunyang Chen}, {and} \bibinfo{person}{Zhenchang Xing}.} \bibinfo{year}{2023}\natexlab{}.
\newblock \showarticletitle{Exploring ai ethics of chatgpt: A diagnostic analysis}.
\newblock \bibinfo{journal}{\emph{arXiv preprint arXiv:2301.12867}} (\bibinfo{year}{2023}).
\newblock


\end{thebibliography}

\appendix

\section{Example of Evaluation Metrics} \label{app: toy_example}

\begin{figure}[h]
    \centering
    \includegraphics[width=1\columnwidth]{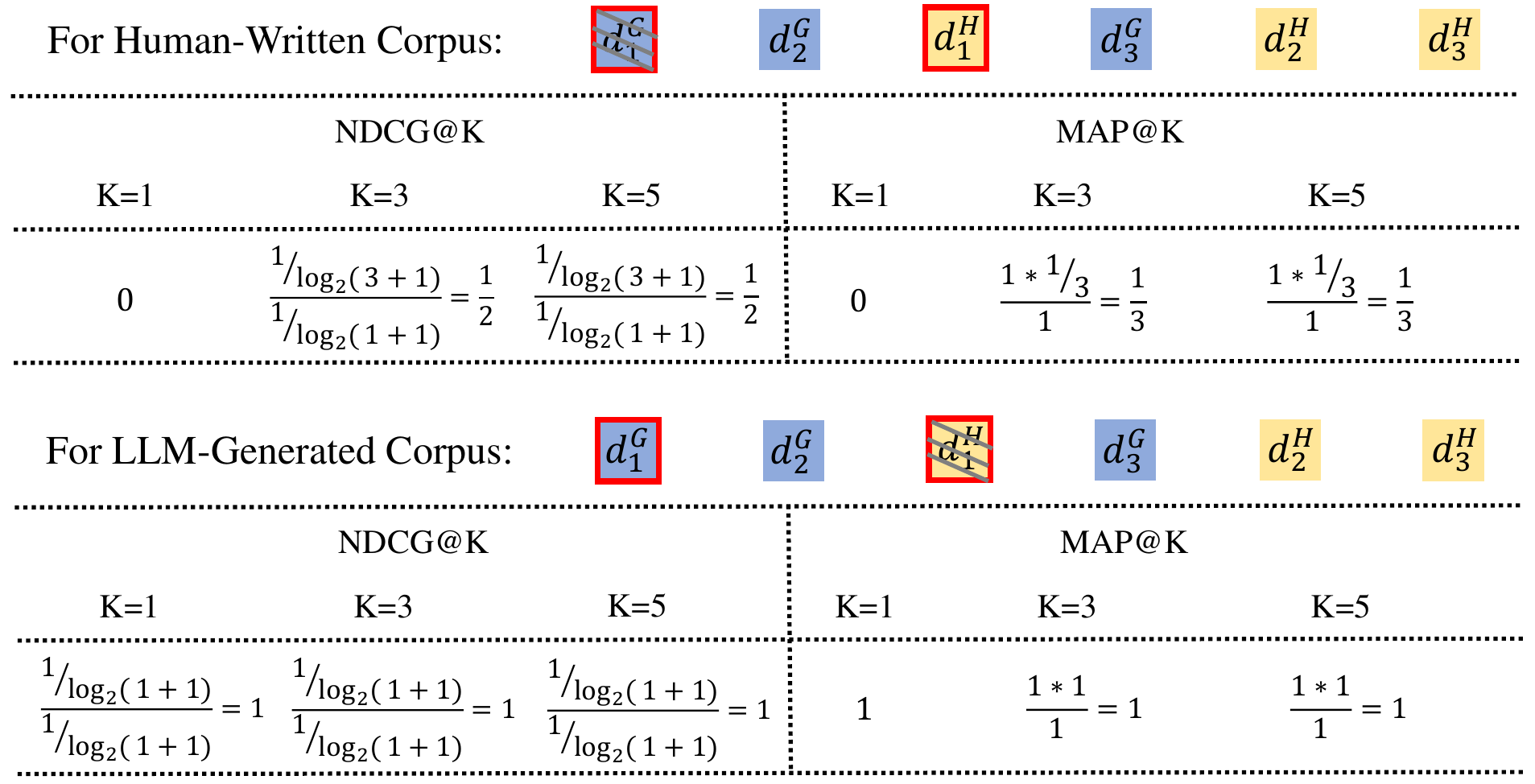}
    \captionof{figure}{A toy example to illustrate how ranking metrics are calculated for each target corpus. In this toy example, given a query, the top 6 documents are retrieved and the rank is from left to right in descending order. Two relevant documents (i.e., $d_1^G$ and $d_1^H$) are highlighted with red boxes.}
    \label{fig:toy_example_metrics}
\end{figure}

In this section, we provide a toy example for illustrating the calculation of the evaluation metrics for exploring source bias, as depicted in \autoref{fig:toy_example_metrics}. 
Specifically, for each query, an IR model produces a ranking list that comprises documents from both human-written and LLM-generated corpora. 
We then calculate ranking metrics separately for human-written and LLM-generated texts, depending on the target data source. When computing metrics for one target (e.g., human-written corpus $\mathcal{C}^{H}$), the data corresponding to the other side (e.g., LLM-generated corpus $\mathcal{C}^{G}$) is disregarded (i.e., mask all the positive label $r \in \mathcal{R}^{G}$ as negative), but the rank of each document is still based on the original ranking list that incorporates a mixture of both types of text. 
For instance, in this toy example, when targeting the human-written corpus, the relevant document $d_1^G$ generated by LLM is treated as a negative sample. And when calculating the ranking metrics, we only consider the rank of the positive human-written documents. When the target corpus is LLM-generated, we adopt the same principle, i.e., only take the rank of the positive LLM-generated documents into account for calculating the ranking metrics.

\section{More Experimental Results} \label{app:more_prompts}

In \autoref{tab:source_bias_ndcg1_prompt}, we provide the results for source bias with more common prompts sourced from InstructGPT-prompts Github Repository~\footnote{\url{https://github.com/kevinamiri/Instructgpt-prompts\#rephrase-a-passage}}. These results indicate that common prompts can easily trigger source bias with LLM-generated content.

\begin{table} [!h]
\caption{Overall source bias in neural retrievers w.r.t. Relative~$\Delta$ (NDCG@1) on SciFact+AIGC with mixed human-written and Llama2-generated corpora generated from different common rephrasing prompts.
}
\label{tab:source_bias_ndcg1_prompt}
\resizebox{1.0\columnwidth}{!}{
\begin{tabular}{ccccccc}
\hline \hline
Prompt & ANCE & BERM & TAS-B & Contriever \\
\hline
Rewrite the text below in your own words: & -7.0 & -22.8 & -11.2 & -27.6 \\
Paraphrase the provided text while maintaining its meaning: & -26.0 & -55.3 & -24.1 & -13.6 \\
Summarize the following passage in a concise manner: & -1.4 & -43.3 & -34.0 & -32.4 \\
Simplify the given passage while keeping the main ideas: & -29.0 & -21.7 & -22.5 & -40.9 \\
Rephrase the given text using alternative expressions: & -25.3 & -34.7 & -61.9 & -18.4 \\
Condense the following passage to focus on key points: & -19.0 & -24.8 & -22.8 & -16.4 \\
Briefly restate the provided text without losing its essence: & -29.6 & -50.7 & -41.3 & -29.8 \\
Reword the passage below to make it more succinct: & -40.6 & -71.1 & -54.7 & -34.0 \\
Express the following text in a different way while keeping its intent: & -50.7 & -39.8 & -34.9 & 0.0  \\
\hline \hline
\end{tabular}
}
\end{table}

\section{Theoretical Analysis and Insights} \label{app: theoretical_ana}
In ~\autoref{fig:scifact_ppl}, we have compared the PPL for different corpus using the BERT model. 
In this section, we aim to further provide some theoretical insights into the above observations that LLM-generated texts have a smaller perplexity than human-written texts.

Without loss of generality, we define the PPL in an autoregressive manner.
Let $d^H$ denote a document written by humans, and $d^G$ a document generated by an LLM conditioned on $d^H.$ For a given document $d$ and BERT model $\mathcal{B}$, PPL is calculated as 
\[\text{PPL}(d, \mathcal{B}) = -\frac{1}{S} \sum_{s=1}^{S} \log P_\text{BERT}(d_s|d_{< {s}}).\]
Similarly, we use $\text{PPL}(d, \mathcal{H})$ to represent the PPL of document $d$ when evaluated by humans. The PPL of $d^G$ conditioned on $d^H$ is denoted as 
$$ \text{PPL}(d^G \mid d^H, \mathcal{B}) = -\frac{1}{S} \sum_{s=1}^{S} \log P_\text{BERT}(d_s^G|d_{< {s}}^G, d^H).$$ 
When evaluated by humans, we use $\text{PPL}(d^G \mid d^H, \mathcal{H})$ to represent the PPL of $d^G$ conditioned on $d^H$. 

In the theorem below, we introduce three assumptions: Semantic Superiority, Conditional Redundancy, and Bounded Perplexity, to theoretically establish the sufficient conditions under which $ \text{PPL}(d^G, \mathcal{B}) \leq \text{PPL}(d^H, \mathcal{B})$ holds. Semantic Superiority suggests that the perplexity of human-written texts, when evaluated by humans, is lower than when evaluated by BERT. Conditional Redundancy implies that the perplexity of $d^G$, given $d^H$, is lower than the perplexity of $d^H$ when evaluated directly. This is intuitively true when the information added in generating from \(d^H\) to \(d^G\) doesn't exceed the original information in \(d^H\). 
Bounded perplexity assumes that there exists an upper bound $\epsilon$ on the increase in perplexity when evaluating $d^G $directly, compared to evaluating $d^G$ conditioned on $d^H$. 
Then we have the following theorem:

\begin{theorem} \label{theo: ppl}
Given the following conditions:
\begin{itemize}  [leftmargin=0.5cm]
    \item Semantic Superiority: human beings outperform BERT in understanding human-written texts, i.e., 
		$$ \textnormal{PPL}(d^H, \mathcal{B}) - \textnormal{PPL}(d^H, \mathcal{H}) \geq 0.$$
    \item Conditional Redundancy: generating \(d^G\) from \(d^H\) adds less perplexity than \(d^H\) itself, i.e., 
    $$ \textnormal{PPL}(d^H, \mathcal{H}) -\textnormal{PPL}(d^G \mid d^H, \mathcal{H})\geq 0.$$

    \item Bounded Perplexity: there exists a bounded non-negative difference $ \epsilon$ in BERT's perplexity for $d^G$ with or without $d^H,$ i.e.,
    $$\textnormal{PPL}(d^G, \mathcal{B})- \textnormal{PPL}(d^G \mid d^H, \mathcal{B}) \leq \epsilon.$$
\end{itemize}
    If LLM aligns more closely with BERT than with humans when predicting $d^G$ given $d^H$, such that for any $s \in [S],$ 
    \begin{equation}
        \begin{aligned}\label{eqn:KL-distance}
            &D_{\textnormal{KL}} \left(P_\text{LLM}(d_s^G|d_{< {s}}^G, d^H) \| P_\text{BERT}(d_s^G|d_{< {s}}^G, d^H)\right) + \epsilon \\
        &\leq  D_{\textnormal{KL}}\left(P_\text{LLM}(d_s^G|d_{< {s}}^G, d^H) \| P_\text{Human}(d_s^G|d_{< {s}}^G, d^H) \right),
        \end{aligned}
    \end{equation} it follows that
    $$
    \mathbb{E}_{P_{\text{LLM}}(d^G \mid d^H)}\left[\textnormal{PPL}(d^G, \mathcal{B}) - \textnormal{PPL}(d^H, \mathcal{B})\right] \leq 0.
    $$
\end{theorem}

In ~\autoref{theo: ppl}, the KL divergence is used to compare the distributions of the document $d^G$ conditioned on $d^H$ according to the LLM, BERT model, and humans. It is worth emphasizing that inequation ~\eqref{eqn:KL-distance} is not the assumption on the understanding capabilities of BERT, LLM, and humans. Instead, this inequation assumes that when predicting $d^G$ given $d^H,$ the predictions by LLM are more closely aligned with those of BERT. 

We demonstrate that, when inequation ~\eqref{eqn:KL-distance} is satisfied, the perplexity (evaluated by PLMs such as BERT) of $d^G$ is lower than that of $d^H$. We'd like to emphasize that it is reasonable to expect that inequation ~\eqref{eqn:KL-distance} holds true because both LLM and BERT are Transformer-based models that use similar pretraining paradigms. The commonality in model structure and learning paradigms may lead to similar inherent biases in text prediction, making their predictions more aligned with each other.

The proof for ~\autoref{theo: ppl} is provided as follows:

\begin{proof}
We start by introducing the term $ \text{PPL}(d^H, \mathcal{H}):$
	\begin{align*}
		\text{PPL}&(d^G, \mathcal{B}) - \text{PPL}(d^H, \mathcal{B}) \\
		&= \text{PPL}(d^G, \mathcal{B}) - \text{PPL}(d^H, \mathcal{H}) + \text{PPL}(d^H, \mathcal{H}) - \text{PPL}(d^H, \mathcal{B}) \\
		&\leq \text{PPL}(d^G, \mathcal{B}) - \text{PPL}(d^H, \mathcal{H}),
	\end{align*}
	where the last step follows from the Semantic Superiority condition.
	\begin{align*}
		\text{PPL}& (d^G, \mathcal{B}) - \text{PPL}(d^H, \mathcal{B}) 
		\leq \text{PPL}(d^G, \mathcal{B}) - \text{PPL}(d^H, \mathcal{H}) \\
		& = \text{PPL}(d^G, \mathcal{B}) - \text{PPL}(d^G \mid d^H, \mathcal{G}) \\
         & \quad + \text{PPL}(d^G \mid d^H, \mathcal{G})  - \text{PPL}(d^H, \mathcal{H}).
	\end{align*}
	Next, we provide upper bounds of $\text{PPL}(d^G \mid d^H, \mathcal{G})  - \text{PPL}(d^H, \mathcal{H}):$
	\begin{align*}
		&\text{PPL}(d^G \mid d^H, \mathcal{G}) - \text{PPL}(d^H, \mathcal{H}) \\
      &=\text{PPL}(d^G \mid d^H, \mathcal{G}) - \text{PPL}(d^G \mid d^H, \mathcal{H}) \\
      & \quad + \text{PPL}(d^G \mid d^H, \mathcal{H}) - \text{PPL}(d^H, \mathcal{H}) \\
		& \leq \text{PPL}(d^G \mid d^H, \mathcal{G}) - \text{PPL}(d^G \mid d^H, \mathcal{H}),
	\end{align*}
	where the inequality follows from the Conditional Redundancy.  
	Taking expectation on $\text{PPL}(d^G \mid d^H, \mathcal{G}) - \text{PPL}(d^G \mid d^H, \mathcal{H}):$
	\begin{align*}
		& -S\mathbb{E}_{P_{\text{LLM}}(d^G \mid d^H)}\left[\text{PPL}(d^G \mid d^H, \mathcal{G}) - \text{PPL}(d^G \mid d^H, \mathcal{H})\right] \\
		& =  \sum_{s=1}^{S} \mathbb{E}_{P_{\text{LLM}}(d^G \mid d^H)} \log \frac{P_\text{LLM}(d_s^G|d_{< {s}}^G, d^H)}{P_\text{Human}(d_s^G|d_{< {s}}^G, d^H)} \\
		& =  \sum_{s=1}^{S} \mathbb{E}_{P_{\text{LLM}}(d_{<s}^G \mid d^H)} \mathbb{E}_{P_{\text{LLM}}(d_s^G \mid d_{<s}^G, d^H)} \log \frac{P_\text{LLM}(d_s^G|d_{< {s}}^G, d^H)}{P_\text{Human}(d_s^G|d_{< {s}}^G, d^H)} \\
		& =  \sum_{s=1}^{S} \mathbb{E}_{P_{\text{LLM}}(d_{<s}^G \mid d^H)} 
        \underbrace{D_{\textnormal{KL}}(P_\text{LLM}(d_s^G|d_{< {s}}^G, d^H) \| P_\text{Human}(d_s^G|d_{< {s}}^G, d^H) )}_{D_{\textnormal{KL}}(P_\text{LLM} \| P_\text{Human})}.
	\end{align*}
	Similarly, $\text{PPL}(d^G, \mathcal{B}) - \text{PPL}(d^G \mid d^H, \mathcal{G})$ can be rewritten as:
	\begin{align*}
		&\text{PPL}(d^G, \mathcal{B}) - \text{PPL}(d^G \mid d^H, \mathcal{G}) \\ 
       &=\text{PPL}(d^G, \mathcal{B})- \text{PPL}(d^G \mid d^H, \mathcal{B}) \\
       & \quad + \text{PPL}(d^G \mid d^H, \mathcal{B}) - \text{PPL}(d^G \mid d^H, \mathcal{G}).
	\end{align*}
	Taking expectation on $\text{PPL}(d^G \mid d^H, \mathcal{B}) - \text{PPL}(d^G \mid d^H, \mathcal{G}):$
	\begin{align*}
		&S\mathbb{E}_{P_{\text{LLM}}(d^G \mid d^H)}\left[\text{PPL}(d^G \mid d^H, \mathcal{B}) - \text{PPL}(d^G \mid d^H, \mathcal{G})\right] \\
		& = \sum_{s=1}^{S} \mathbb{E}_{P_{\text{LLM}}(d_{<s}^L \mid d^H)}
        \underbrace{D_{\textnormal{KL}}(P_\text{LLM}(d_s^G|d_{< {s}}^G, d^H) \| P_\text{BERT}(d_s^G|d_{< {s}}^G, d^H) )}_{D_{\textnormal{KL}}(P_\text{LLM} \| P_\text{BERT})}.
	\end{align*}
	Thus, 
	\begin{align*}
		&\mathbb{E}_{P_{\text{LLM}}(d^G \mid d^H)}\left[\text{PPL} (d^G, \mathcal{B}) -\text{PPL} (d^H, \mathcal{B}) \right] \\
		&\leq \mathbb{E}_{P_{\text{LLM}}(d^G \mid d^H)}\left[\text{PPL}(d^G, \mathcal{B})- \text{PPL}(d^G \mid d^H, \mathcal{B}) \right]  \\
		& + \frac{1}{S}\sum_{s=1}^S \mathbb{E}_{P_{\text{LLM}}(d_{<s}^G \mid d^H)} ( D_{\textnormal{KL}}(P_{\text{LLM}} \| P_{\text{BERT}}) - D_{\textnormal{KL}}(P_{\text{LLM}} \| P_{\text{Human}})).
	\end{align*} 
	The final results can be derived by considering the assumptions: 
	\begin{align*}
		\text{PPL}(d^G, \mathcal{B})- \text{PPL}(d^G \mid d^H, \mathcal{B}) &\leq \epsilon \\
        D_{\textnormal{KL}}(P_{\text{LLM}} \| P_{\text{BERT}}) -
		D_{\textnormal{KL}}(P_{\text{LLM}} \| P_{\text{Human}}) &\leq - \epsilon.
	\end{align*}
    From these, it follows that:
    $$
      \mathbb{E}_{P_{\text{LLM}}(d^G \mid d^H)}\left[\text{PPL} (d^G, \mathcal{B}) -\text{PPL} (d^H, \mathcal{B}) \right] \leq \epsilon - \epsilon = 0.
    $$
\end{proof}

\end{document}